\documentclass[12pt,preprint,aps,amssymb,amsmath,floatfix]{revtex4}
\usepackage{epsfig}
\usepackage{graphicx}

\newcommand{\beq}{\begin{equation}}
\newcommand{\eeq}{\end{equation}}
\newcommand{\beqar}[1]{\begin{eqnarray}\label{#1}}
\newcommand{\eeqar}{\end{eqnarray}}

\newcommand{\NJ}{N_{\J}}
\newcommand{\J}{J/\psi}

\newcommand{\Nch}{N_{ch}}

\newcommand{\ccbar}{c \bar c}
\newcommand{\Nccbar}{N_{\ccbar}}
\newcommand{\kts}{\langle{k_T}^2\rangle}
\newcommand{\pts}{\langle{p_T}^2\rangle}
\newcommand{\epsz}{\epsilon_0}
\newcommand{\sigF}{\sigma_F}
\newcommand{\sigD}{\sigma_D}
\newcommand{\gev}{\mathrm{GeV}}

\begin{document}

\title {Momentum spectra of charmonium produced
in a quark-gluon plasma
}

\author{R. L. Thews}

\affiliation{ Department of Physics, University of Arizona, Tucson, AZ 85721 USA}

\author{M. L. Mangano}

\affiliation{PH Department, TH Unit, CERN, 1211 Geneva 23, Switzerland}

\date{\today}

\begin{abstract}

We calculate rapidity and transverse momentum distributions  
of charmonium formed in high energy
heavy ion collisions from incoherent recombination of charm quarks.
The results are very sensitive to the corresponding distributions 
of the charm quarks, and thus can serve as a probe of the state
of matter produced in the heavy ion collision. 
At one extreme we generate a set of charm pair momenta directly from
pQCD amplitudes, which are appropriate if one can neglect interaction
of the quarks with the medium.  At the other extreme we 
generate momenta of charm quarks in  thermal equilibrium
with the expanding medium, appropriate for an extremely strong
interaction.  Explicit predictions are made for $\J$ formation
in Au-Au interactions at RHIC.  We find that for the case in which
charm quark momenta are unchanged from the pQCD production calculation,
both the rapidity and transverse
momentum spectra of the formed $\J$ are 
substantially narrower than would be anticipated in scenarios  which
do not include the in-medium formation.  In particular, the average
transverse momentum of the $\J$ will exhibit a non-monotonic behavior
in the progression from p-p to p-A to A-A interactions.

\end{abstract}

\maketitle


\section{Introduction}
The behavior of heavy quarkonium states as an experimental signature
of medium modification in high energy heavy ion collisions originated
with the 
prediction of $\J$ suppression \cite{Matsui:1986dk}, which follows from 
screening of the color-confinement 
potential above the phase transition temperature.
The suppression argument requires that the probability for
 recombination of the c and $\bar{c}$ quark is negligible.
Recently it has been realized that for RHIC and LHC collider
energies, there can be a modification of the suppression argument
 \cite{Braun-Munzinger:2000px,Thews:2000rj}.  

Consider a central heavy ion collision in which multiple $\ccbar$ 
pairs are produced in binary nucleon-nucleon interactions.  If these
quarks are then immersed in a medium (such as a quark-gluon plasma)
in which a charm quark from one initial $\ccbar$ pair can readily interact
with an anticharm quark from a different initial $\ccbar$ pair, one
expects that the number of $\J$ formed from such interactions will
be proportional to the total number of $\ccbar$ combinations.  The 
relative efficiency of this quadratic process must be normalized
by the number of interactions involving charm plus light-flavor quarks in
which open-charm hadrons are formed.  The normalization measure is
generally taken 
\cite{Thews:2001hy} to be proportional to the total number of charged
hadrons produced in the heavy ion collision, $\Nch$, which leads
to 
$\NJ \propto {\Nccbar}^2 / \Nch.$
(Of course this behavior
must saturate before $\NJ$ becomes comparable in magnitude to $\Nccbar$.)
Then at sufficiently large $\Nccbar$ this
quadratic behavior must dominate over the color-screening suppression.

Initial estimates \cite{Gavai:1994in} 
from extrapolations of the low
energy charm production measurements and predictions of pQCD
indicated of $\Nccbar \approx$ 10 for central RHIC collisions and
several hundred at LHC.  Measurements at RHIC of high transverse
momentum leptons 
in Au-Au collisions by PHENIX \cite{Adler:2004ta} imply that $\Nccbar
\approx $ 20, and measurement of reconstructed D mesons 
in d-Au collisions by 
STAR \cite{Adams:2004fc} require $\Nccbar \approx$ 40.  Statistical and
systematic uncertainties are large enough such that these two
measurements are consistent, but the general trend indicates a
more rapid growth with energy than initially estimated \cite{Vogt:2004he}.

Two distinct realizations of this mechanism for heavy quarkonium
formation  have been developed.  In the statistical hadronization model,
quarks are distributed into hadrons during the hadronization transition
according to chemical equilibrium ratios.  The total number of heavy
quarks is assumed to be determined by initial production, and an extra fugacity
factor $\gamma_f$ is determined by overall flavor conservation. In the
case of hidden flavor such as $\J$, two powers of $\gamma_c$ modify
the thermal equilibrium density, which leads to the expected quadratic
dependence on $\Nccbar$ \cite{Andronic:2003zv,Kostyuk:2003kt}.  
There is no explicit prediction for the
$\J$ momentum spectrum, but it is natural to assume a thermal distribution
in this model at the deconfinement temperature $T_c \approx$ 170 MeV.
It is noteworthy that ratios involving radially-excited states such
as $\frac{\psi^{\prime}}{ \J}$ retain their unmodified statistical values.

The kinetic formation model 
\cite{Thews:2000rj,Thews:2001hy,Schroedter:2000ek,Thews:2000fd,Thews:2001em}
 assumes that heavy quarkonium formation
takes place during the entire lifetime of a color-deconfined phase.
Predictions for the resulting population require the specification
of cross sections for formation and dissociation in the medium. 
The motivation for such a scenario in the case of
$\J$ formation has received support
from recent lattice calculations of spectral functions.  These indicate that
the $\J$ and $\eta_c$ will exist in a thermal environment at temperatures 
well above the deconfinement
transition \cite{Datta:2003ww,Asakawa:2003re}.  
Note that this property is likely valid only for 
the most deeply bound states. Thus, unlike the statistical model, 
one does not expect a simple relation to exist between the
$\J$ and $\psi^{\prime}$ population.
Also in contrast with statistical hadronization, the momentum
spectrum of formed $\J$  will reflect the initially-produced
charm quark spectra 
plus any modification due to interaction with medium.

In principle, both of these formation processes could occur
sequentially. Considerable progress has been made 
\cite{Grandchamp:2001pf,Grandchamp:2002wp,Grandchamp:2003uw}
in such
a scenario when both bound and unbound charm
populations evolve in a thermal fireball.

In the next section we review the basic properties of kinetic 
in-medium formation of heavy quarkonium, using a specific model 
for $\J$ which exhibits the expected general properties.  We have
previously found \cite{Thews:2003da} 
that the initial PHENIX data from RHIC is able to constrain the model
parameters within a fairly broad range, but is unable to
confirm or rule out this type of formation.  A more detailed test must await
comparison with the momentum spectra of the formed $\J$.  Section
\ref {charmquarkdistributions} contains details of the two specific
charm quark distributions which we consider.  One is calculated directly
from the initial production process using NLO pQCD amplitudes to generate
a set of $\ccbar$ pair events appropriate for RHIC energy.  The other
generates a corresponding set of events which would be expected if
the charm quarks subsequently come into local thermal equilibrium with
the expanding medium before forming the $\J$. In Section \ref{kineticpredictions}
we present a detailed model calculation of the {\it normalized} 
rapidity and transverse momentum spectra
for in-medium $\J$ formation, using charm quark spectra calculated from
the pQCD amplitudes.
It is important to keep in mind that we are searching for
a signature in the $\J$ rapidity and transverse momentum spectra which
would indicate the presence of in-medium formation, and which is 
independent of the absolute magnitude of the formation process.
Our results are shown to be quite robust with respect
to substantial variations in the model parameters. Section \ref{pqcd}
contains a specific for RHIC results, where we use data for pp and dAu 
interactions to constrain some of the model parameters.  Section \ref{thermal}
presents the corresponding calculations using charm quarks in local
thermal equilibrium.  Comparisons with predictions of other
models and a summary of results complete this presentation.

\section{Formation of $\J$ in a color-deconfined medium}
\label{formationreview}
We first examine the net number of $\J$ produced in a color-deconfined
medium due to the competing reactions of (a) formation involving recombination
of $c$ and $\bar{c}$ and (b) dissociation of $\J$ induced by interactions
with constituents of the medium.  The simplest dissociation reaction
utilizes absorption of single deconfined gluons in the medium to ionize the 
color singlet $\J$, $g + \J \rightarrow c + \bar{c}$, 
resulting in a $\ccbar$ pair in a color octet state.  This process
was originally proposed \cite{Kharzeev:1994pz} as a dynamical counterpart of the 
static color screening effect.  The inverse of this process then serves
as the corresponding formation reaction, in which a $\ccbar$ pair in
a color octet state emits a color octet gluon and falls into the
color singlet $\J$ bound state. 

One can then follow the time evolution 
of charm quark and charmonium numbers according
to a Boltzmann equation in which the formation and dissociation 
reactions compete.

\begin{equation}\label{eqkin}
\frac{d\NJ}{dt}=
  \lambda_{\mathrm{F}} N_c\, N_{\bar c }[V(t)]^{-1} -
    \lambda_{\mathrm{D}} \NJ\, \rho_g\,,
\end{equation}
with $\rho_g$ the number density of gluons in the medium.
The reactivity $\lambda$ is the product of the reaction
cross section and initial relative velocity 
$\langle \sigma v_{\mathrm{rel}} \rangle$
averaged over the momentum distribution of the initial
participants, i.e. $c$ and $\bar c$ for $\lambda_F$ and
$\J$ and $g$ for $\lambda_D$.
The gluon density is determined by the equilibrium value in the
medium at each temperature, and the volume must be modeled according to the
expansion and cooling profiles of the heavy ion interaction region.

This equation has an analytic solution in the case where the total number
of formed $\J$ is much smaller than the initial number of $\Nccbar$.

\begin{equation}
\NJ(t_f) = \epsilon(t_f) [\NJ(t_0) +
\Nccbar^2 \int_{t_0}^{t_f}
{\lambda_{\mathrm{F}}\, [V(t)\, \epsilon(t)]^{-1}\, dt}],
\label{eqbeta}
\end{equation}
where $t_0$ and $t_f$ define the lifetime of the deconfined region.
Note that the function $\epsilon(t_f) = 
e^{-\int_{t_0}^{t_f}{\lambda_{\mathrm{D}}\, \rho_g\,
dt}}$
would be the suppression factor in this scenario if the
formation mechanism were neglected.

The initial calculations \cite{Thews:2001hy}
used the ratio of nucleon participants to participant
density calculated in a Glauber model 
to define a transverse area enclosed by the boundary of the
region of color deconfinement.  This is supplemented by longitudinal expansion
starting at an initial time $t_{0}$ = 0.5 fm.   Transverse expansion was
initially neglected, but has been included in 
subsequent calculations \cite{Thews:2003da}.
The expansion was taken to be isentropic, which determines the 
time evolution behavior  of
the temperature $T(t)$.  The initial value $T_{0}$ is 
taken as a parameter, and the
final $T_{f}$ is fixed at the hadronization point. 

\begin{figure}[hbt]
\vskip 1.0truecm
\epsfig{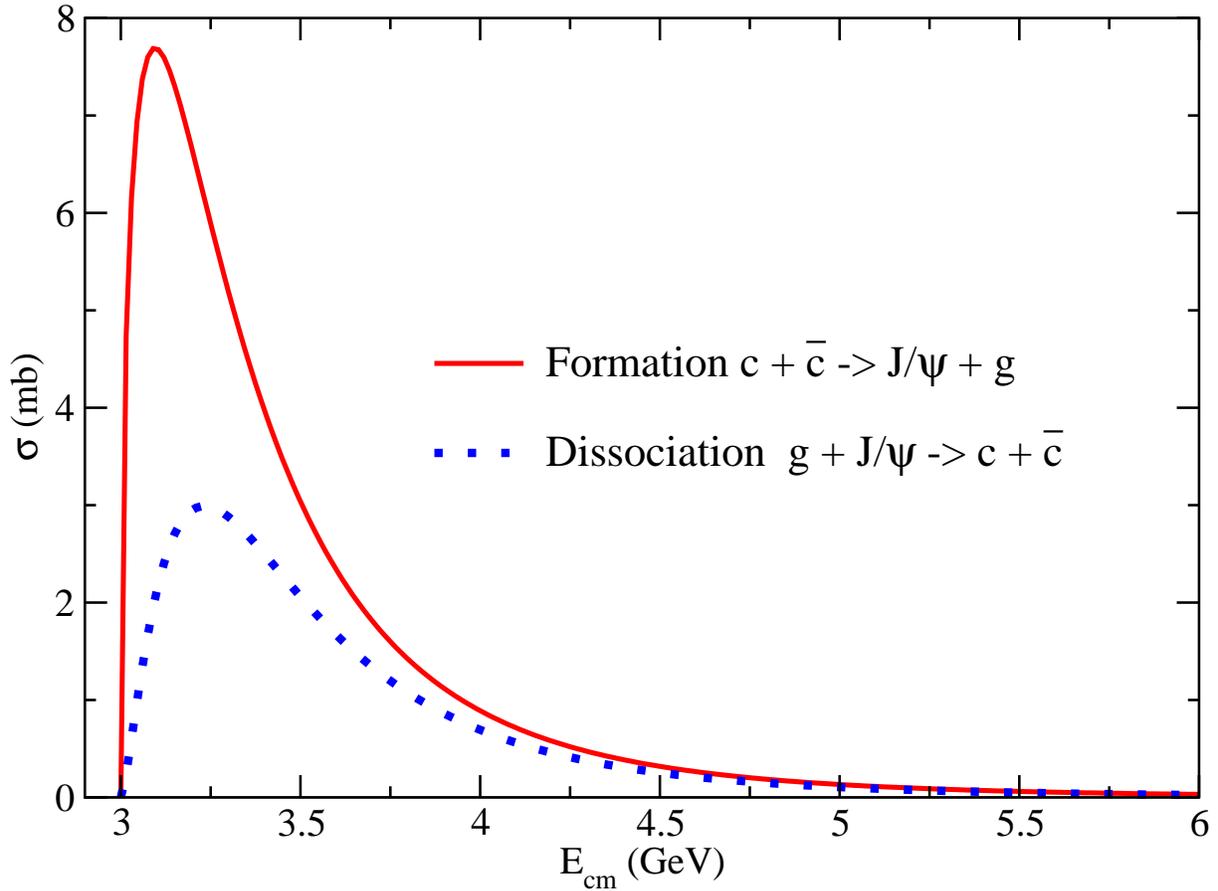}
\caption{(Color online) OPE-based cross sections for $\J$ formation and dissociation.}
\label{opecharmcrosssections}
\end{figure}

The reactivities $\lambda_{F}$
and $\lambda_{D}$ require specification of cross sections.  For $\sigma_{D}$ we
use the OPE-based model of gluon dissociation 
of deeply-bound heavy quarkonium
\cite{Peskin:1979va, Bhanot:1979vb, Kharzeev:1995ij}, which is related via 
detailed balance to the corresponding
$\sigma_{F}$.  Written in terms of the heavy quarkonium reduced mass
$\mu (= m_Q/2)$, gluon momentum in the $\ccbar$ rest frame $k$ and binding energy $\epsz$, the dissociation
cross section is

\begin{equation}
\sigma_{D} = \frac{2\pi}{3} \left (\frac{32}{ 3}\right )^2
\left (\frac{2\mu}{\epsz}\right )^{1/2}
\frac{1}{4\mu^2} \frac{(k/\epsz - 1)^{3/2}}{(k/\epsz)^5},
\label{eqsigma}
\end{equation}
This expression assumes  that the heavy quarkonium has a 
nonrelativistic Coulomb
bound state spectrum with $\epsz >> \Lambda_{QCD}$, and utilizes an
operator product expansion in the large $N_c$ limit.  
These cross sections for $\J$ kinematics are shown in 
Fig.~\ref{opecharmcrosssections}.
One sees that they are peaked at low energy, and that $\sigma_{F} > \sigma_{D}$
due to the large binding energy (we used the vacuum value $\epsz \approx 
600$~MeV).

We should note that the approximations concerning a nonrelativistic
Coulomb bound state spectrum are somewhat marginal for the 
charmonium case, although
they are better-justified for the upsilon states.
Therefore for 
applications to $\J$ we will investigate the effects of a considerable
range of alternative cross section behavior, as has been done in
a number of treatments of $\J$ production in a purely hadronic 
scenario\cite{Xu:2002zv, Martins:1994hd, Navarra:2001jy, Maiani:2004py}.

The initial model calculations for $\J$ at RHIC \cite{Thews:2001hy} 
yielded interesting
predictions for the centrality behavior of the final population 
$\NJ (t_f)$ normalized by $\Nccbar$, or equivalently normalized by
the number of binary collisions.  Due to the quadratic behavior of
the formation rates with $\Nccbar$, one expects an increase with
centrality of formed $\J$ per binary collision,
 if the formation reaction dominates.
The primary variation in these
predictions was due to the initial charm quark momentum distributions, 
which enter through the calculation of the formation and dissociation
reactivities as defined in Eq.~\ref{eqkin}.  Differences of 
up to a factor of 5 or more resulted as the charm quark momentum
distributions varied from thermal (produced the largest formation
efficiency)  through perturbative QCD distributions.  There may
be additional uncertainties in overall magnitude.  In particular,
an implementation of this process in a transport model 
calculation  \cite{Zhang:2002ug}
did not result in a large effect at RHIC energy.

The first 
confrontation with experiment involved the initial PHENIX measurements
of $\J$ at 200 GeV \cite{Adler:2003rc}.
Values of d$\NJ$/dy (y=0) for three centrality bins were extracted, but
the most central bin value only yielded an upper bound due to the
limitations of statistics.  Although these limitations prevented
a definitive test of the formation model, it was clear that the
predictions which followed from thermal charm quark distributions 
were disfavored.  The pQCD charm quark distributions led to
formation rates which were roughly consistent with experiment.  
A parameter scan involving initial temperature,
initial
charm quark production, transverse expansion
velocity of the deconfined region, and initial $\J$
production number found a surprisingly
large region of parameter space within which the formation mechanism
results were consistent with the PHENIX data points and limits.

It is interesting that these data are also consistent with 
binary-scaled $\J$ production in pp interactions \cite{Adler:2003qs},
modified with a small suppression factor. Given the sensitivity of
the formation process to initial charm production (both magnitude and
momentum spectra), even the improved statistics anticipated for the
PHENIX Au-Au data currently undergoing analysis may not be adequate
to differentiate between these distinct possibilities.  Fortunately, 
the momentum spectra for the produced $\J$ will also be measured.  
In anticipation of this information, we then proceed to extend the
calculations of the kinetic formation process to include the rapidity
and transverse momentum spectra of the final $\J$ population.

We return to the expression for in-medium formation of $\J$ in Eq.~\ref{eqkin}, 
modified to calculate the momentum spectra of the formed $\J$.
Rather than specifying a functional form for the charm
quark momentum distributions, we will generate a sample of $\ccbar$
events appropriate for a given physical situation and then sum
over the sample to calculate the differential formation reactivity.

\begin{equation} 
\frac{dN_{\J}}{d^3 P_{\J}} = \int{\frac{dt}{V(t)}}
\sum_{i=1}^{N_c} \sum_{j=1}^{N_{\bar c}} {\it {v}_{rel}} 
\frac{d \sigma}{d^3 P_{\J}}(P_c + P_{\bar{c}} \rightarrow P_{\J} + X)
\label{formdist}
\end{equation}
Note that the formation magnitude exhibits the explicit quadratic
dependence on $\Nccbar$ via the double summation.

The formation rate (differential in $\J$ momentum) is integrated
over time to get the spectra of in-medium formed $\J$.
We assume for simplicity that (a) the spatial density of charm and anticharm
 quarks are equal within the deconfinement region, and (b) the charm quark 
momentum spectra are independent of time. The time-independence assumption 
could in principle be relaxed, at the expense of a more involved
calculational scheme.  For now we simply present the results as
limiting cases which bracket a range of possibilities between the 
two types of charm momentum distributions considered.
Then the 
differential $\J$ spectra just involve an integral of the inverse
deconfinement volume over time, which is irrelevant for the {\it normalized}
momentum spectra.  Of course, these spectra will be modified by the
dissociation process (second term in Eq.~\ref{eqkin}), as well as 
the spectra of $\J$ which are initially-produced. We consider the
effects of the dissociation process in Sec. \ref{kineticpredictions}.

The differential dependence of the formation cross section which 
follows from detailed balance of the dissociation cross section in Eq. 
\ref{eqsigma} is extracted from the the analog of atomic photoionization.
In this case the relevant amplitude involves the coupling of a gluon 
with the $\J$
color electric dipole of the quarkonium, again in the approximation of
a Coulombic bound state wave function.  The procedure to evaluate
the $\J$ momentum spectra from Eq. \ref{formdist} just involves
a scan for each $\ccbar$ over final state phase space, weighted
by the appropriate differential cross section.  To determine the
sensitivity of the resulting $\J$ spectra to this particular 
differential cross section, we have also performed the calculations
using an isotropic invariant amplitude, normalized to the same OPE-based
total cross section.  We find in general that the effects on the
$\J$ momentum spectra are negligible.  Comparison of results will
be shown in Sec. \ref{kineticpredictions}. 
 
\section{Charm quark momentum spectra}
\label{charmquarkdistributions}
\subsection{Perturbative QCD calculations}
The perturbative QCD spectra are obtained using the results of the
full ${\cal {O}}(\alpha_s^3)$ calculation, as implemented in the event
generator described in~\cite{Mangano:1991jk}. In the case of nuclear
beams, the parton
distribution functions of the proton have been modified to account for nuclear
effects, using the model of ref.~\cite{Eskola:1998iy}. 
The terms of ${\cal
  {O}}(\alpha_s^3)$ corresponding to real-emission diagrams will
generate a non-trivial transverse momentum distribution for the
$\ccbar$ pair. We generated samples of unweighted events, which will
be used in the following analyses. To ensure that all events in the
samples have
positive weight, we merged  the negative-weight events arising from
the cancellation of soft and collinear
singularities~\cite{Mangano:1991jk} with the sample
of positive weight events having a transverse momentum of the $\ccbar$
pair $p_T(\ccbar)<0.4$~GeV, and smeared their   $p_T(\ccbar)$ over the
$0-0.4$~GeV range.
$\ccbar$ pairs selected from the
same event will be labeled as ``diagonal'' pairs: they dominate
the $\J$ formation in $pp$ and $pA$ collisions, where production of
multiple pairs is negligible.
In the case of $AA$ collisions, multiple hard scatterings
can give rise to several $\ccbar$ pairs, and charm (anticharm) quarks
from independent pairs are allowed to form a $\J$. These uncorrelated
pairs are randomly selected from our sample of unweighted events,
reproducing in average the correct kinematical distributions of such
pairs.

\begin{figure}[hbt]
\vskip 1.0truecm
\epsfig{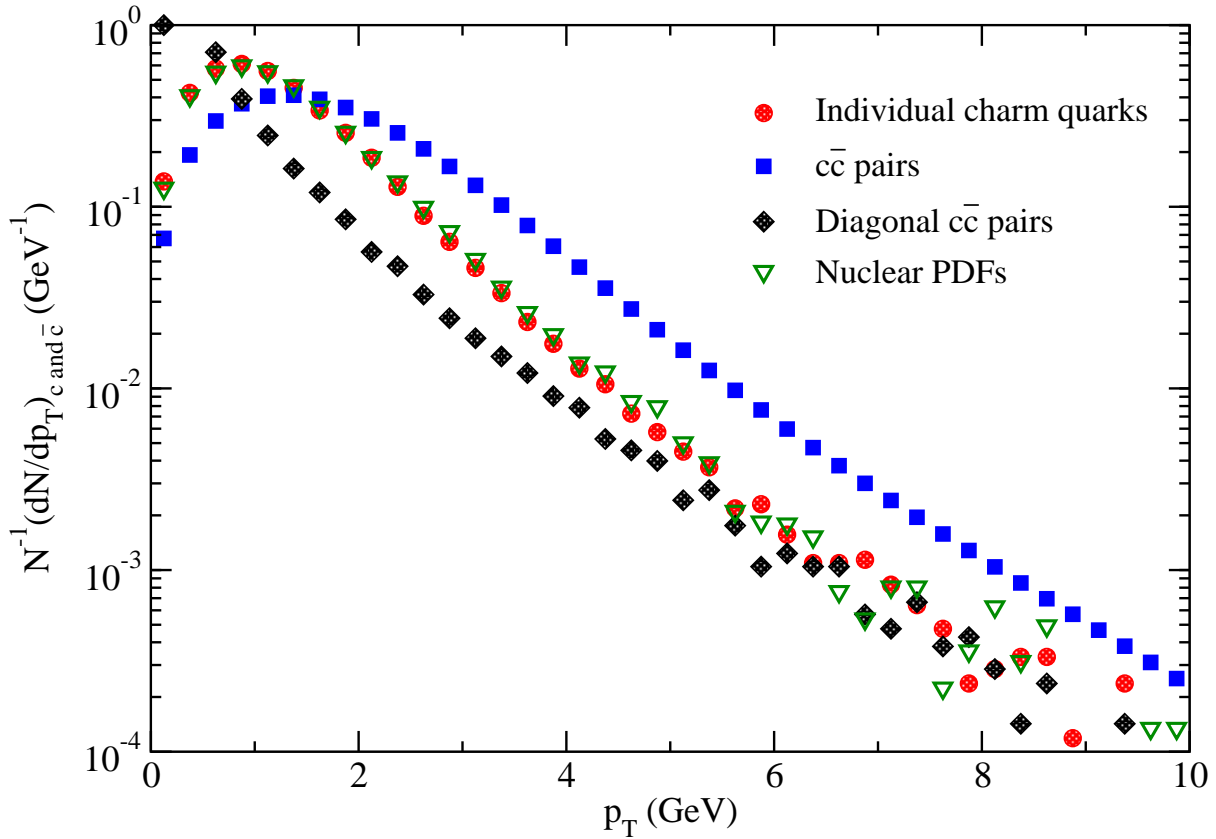}
\caption{(Color online) Transverse momentum distributions from the NLO pQCD $\ccbar$
events at RHIC200 energy.  The individual charm quark distribution
is shown by the
circles.  The $\ccbar$ pair distribution is shown
by the squares, and the subset of diagonal $\ccbar$ pairs is shown by
the diamonds.  (Since we use a sample of some 80K $\ccbar$ events,
the distribution of all pairs is dominated by the off-diagonal pairs.)
Also shown by the open triangles is the individual charm
quark distribution generated using parton distribution
functions modified for nuclear effects by the EKS98 \cite{Eskola:1998iy}
factors.}
\label{cquarkptdist}
\end{figure}

Shown in Fig.~\ref{cquarkptdist} are the shapes of the
transverse momentum distributions
\footnote{A recent review of the
overall theoretical systematics at 
${\cal   {O}}(\alpha_s^3)$ of the charm 
spectrum in the RHIC energy range is given in 
ref.~\cite{Cacciari:2005rk}.}. 

We plot the spectra of individual charm quarks (with and without
inclusion of nuclear effects in the PDFs), as well as the transverse
momenta of $\ccbar$ pairs. Here and in the following the generic term
``$\ccbar$ pairs'' will refer to summing over all possible pairings of
$c$ and $\bar{c}$ quarks in an event, in the limit of large $\Nccbar$
multiplicity. The term ``diagonal pairs'' will refer instead to
$\ccbar$ pairs coming from the same hard collision.

The nuclear effects are essentially negligible in the individual charm
quark distributions (compare the solid circles and open triangles - the
scatter at large $p_t$ reflects statistical fluctuations due to our
finite sample size).  The $\ccbar$ pair distribution is approximately
twice as wide as that for individual quarks (as measured by $\pts$).
This is a result of the independence of the transverse momentum
vectors (both in size and in azimuthal direction) for off-diagonal
pairs of $c$ and
$\bar{c}$ quarks produced in different hard scatterings. For diagonal pairs the
relative azimuthal angle is $\pi$ for all LO events.  The small
nonzero $\pts$ for diagonal pairs thus 
arises entirely from NLO effects. 
The mean values $\langle{p_T}^2\rangle$
of these distributions are 2.45 $\gev^2$ for the individual quarks,
4.88 $\gev^2$ for all pairs, and 1.20 $\gev^2$ for the diagonal pairs.

\begin{figure}[hbt]
\vskip 1.0truecm
\epsfig{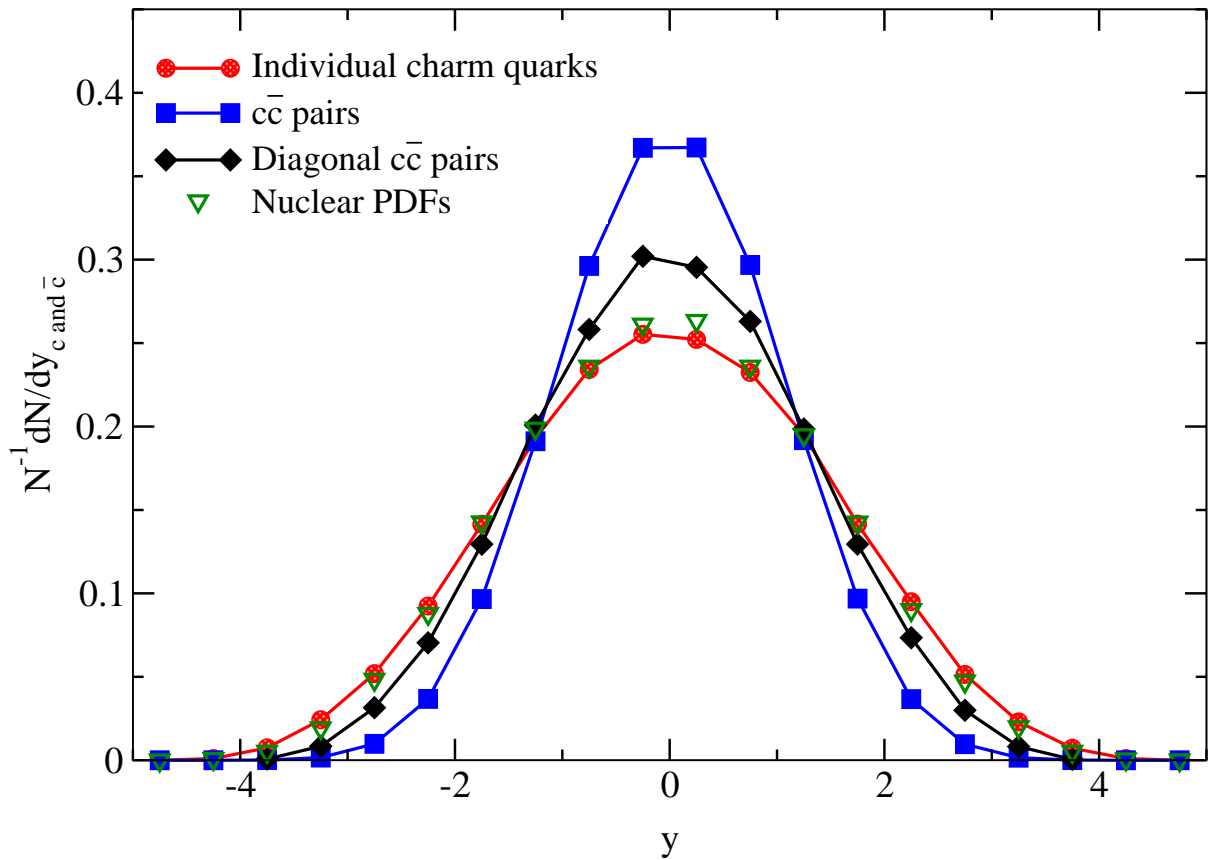}
\caption{(Color online) Rapidity Distributions from the NLO pQCD $\ccbar$ events at
RHIC200 energy. Legends are the same as in Fig. \ref{cquarkptdist}.}
\label{cquarkydist}
\end{figure}

Shown in Fig.~\ref{cquarkydist} are the corresponding rapidity distributions.
Again the nuclear effects are negligible in the individual charm quark
distributions.
The $\ccbar$ pair distributions
are both narrower than that for individual quarks.  
The diagonal pair distribution is somewhat
wider that that of all pairs, which again follows from the kinematics of
diagonal pairs which have $p_t(pair) \approx 0$ from the LO events.  These
widths can be quantified by the $\Delta y \equiv \sqrt{\langle
  y^2\rangle-\langle y\rangle^2}$ values,
1.46 for individual charm quarks, 1.25 for diagonal pairs, and
1.03 for all pairs. 

\subsection{Intrinsic $k_T$ effects}
We must now modify the charm quark transverse momentum spectra to include
effects of partonic confinement. This could be accomplished by introducing
partonic transverse momentum dependence directly into the pQCD calculation.
For simplicity, we choose to modify the NLO charm quark results by
adding an additional transverse momentum `kick' $k_T$ to each quark in a 
$\ccbar$ pair~\cite{Mangano:1992kq}. 
 The magnitude is determined by a Gaussian distribution
with width parameter $\kts$, and uniform azimuthal distribution.  
We 
will determine the width parameter directly from data.  For nuclear 
collisions, the additional effects of initial state interactions 
of the nucleons will be included in this parameterization.

\begin{figure}[hbt]
\vskip 1.0truecm
\epsfig{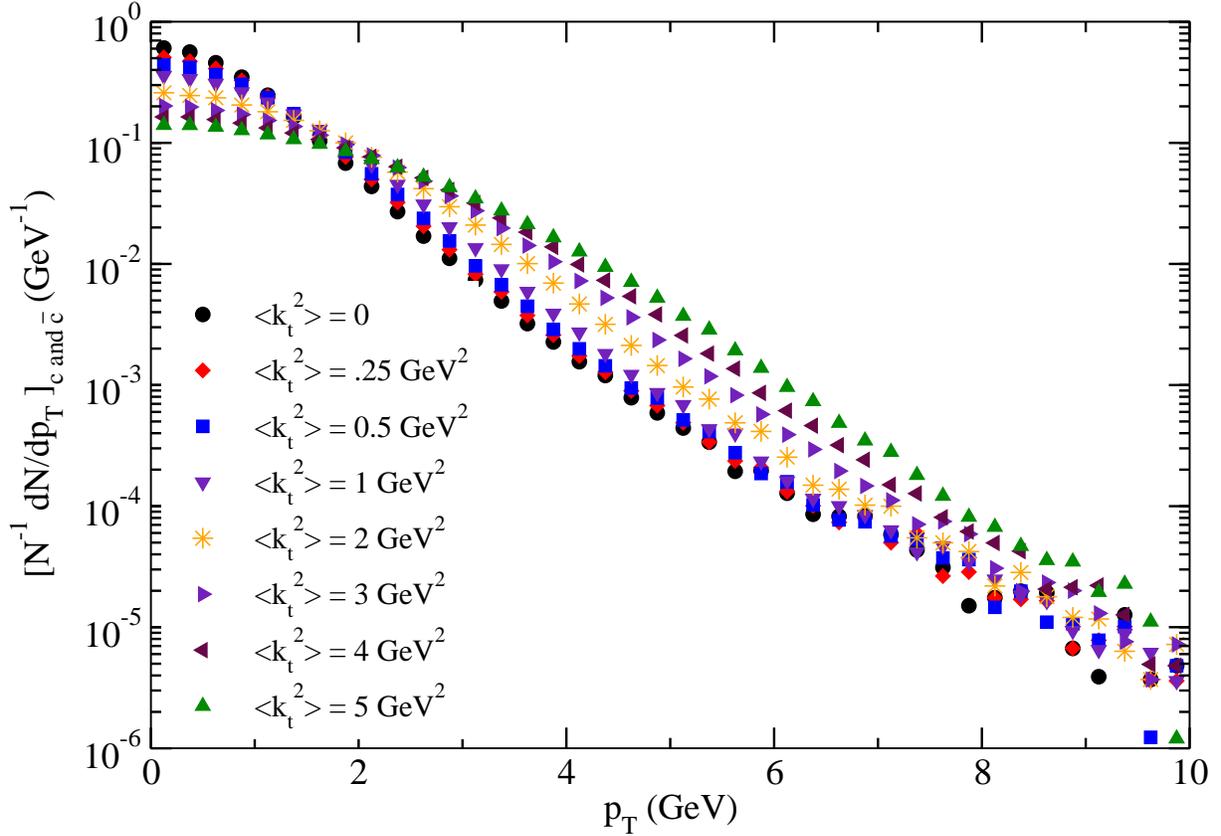}
\caption{(Color online) Charm quark transverse momentum distributions including intrinsic
$k_T$.}
\label{cquarkplusktdist}
\end{figure}

The resulting charm quark transverse momentum spectra are shown in
Fig.~\ref{cquarkplusktdist} for a range of $\kts$.  The width of these
spectra satisfy a simple pattern.
\begin{equation}
\langle{p_T}^2\rangle\left(\kts\right) = \langle{p_T}^2\rangle\left(\kts=0\right) + \kts 
\end{equation}
This pattern is again a result of the uncorrelated relative azimuthal
angles between the  
transverse momentum kick and the transverse momentum of the original
pQCD-generated quarks.

\begin{figure}[hbt]
\vskip 1.0truecm
\epsfig{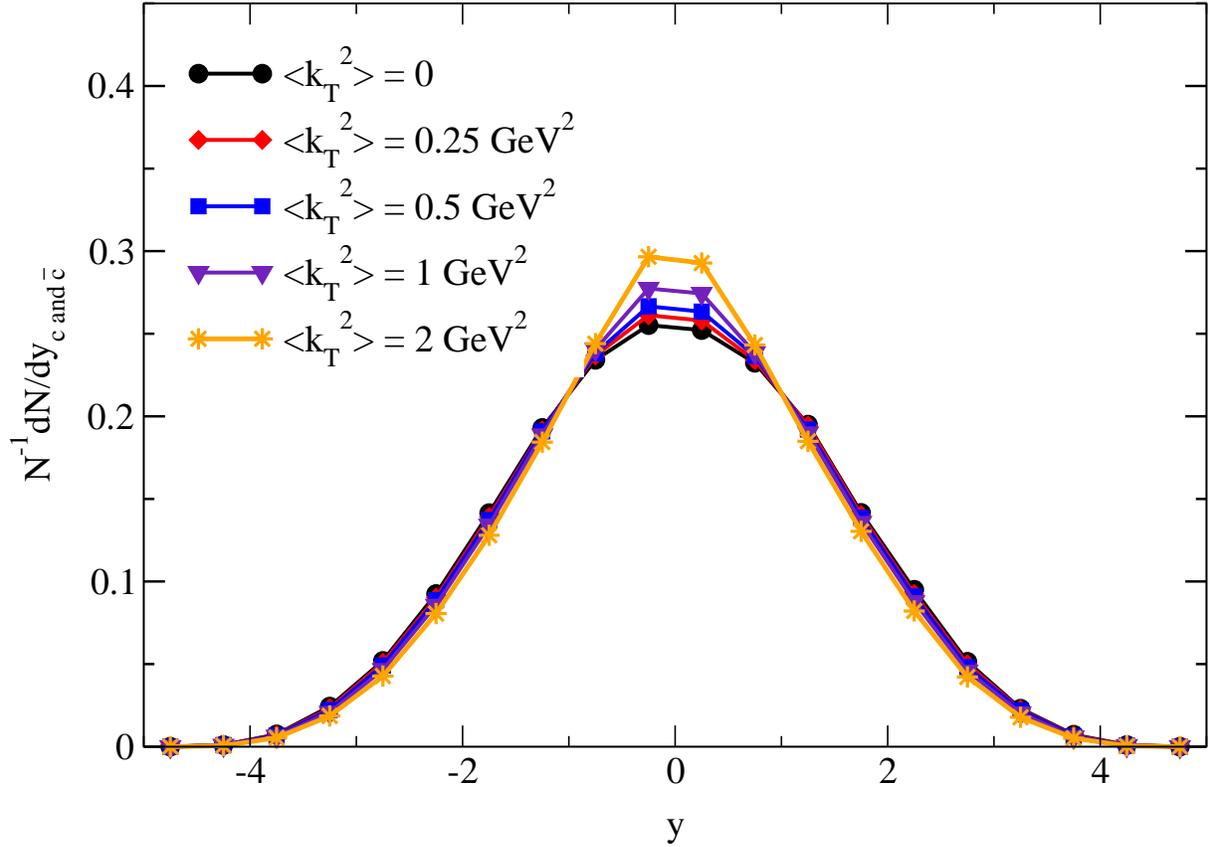}
\caption{(Color online) Charm quark rapidity distributions including intrinsic
$k_T$.}
\label{cquarkplusktrapiditydist}
\end{figure}

Fig.~\ref{cquarkplusktrapiditydist} shows the rapidity distributions of
the charm quark events after the $\kts$ modification.  Aside from small
corrections near midrapidity, these spectra are effectively independent
of $\kts$, because the charm quark longitudinal momenta are unchanged
by addition of the transverse `kick'. 

\begin{figure}[hbt]
\vskip 1.0truecm
\epsfig{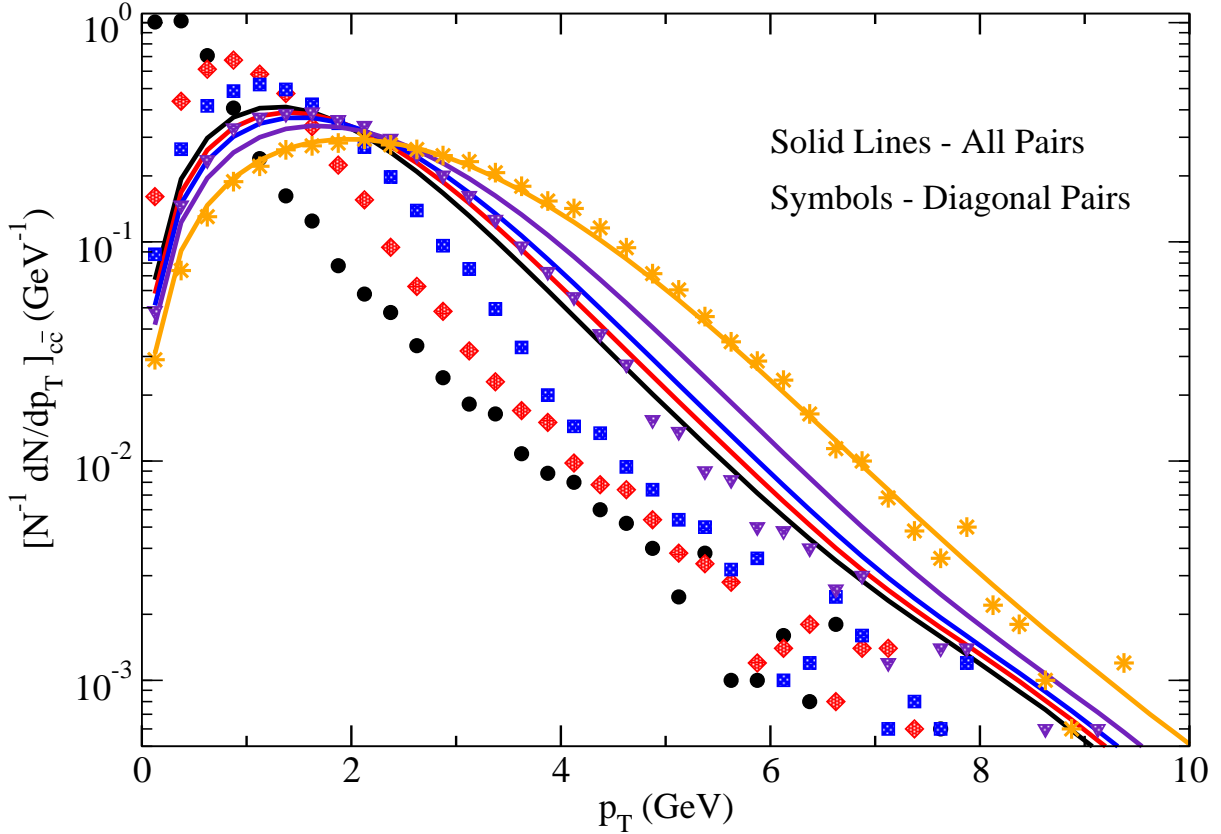}
\caption{(Color online) $\ccbar$ pair transverse momentum distributions including
intrinsic
$k_T$.  Shown are all pairs (solid lines) and diagonal pairs (symbols,
as in Fig.~\ref{cquarkplusktrapiditydist}).
}
\label{ccbarpairplusktdist}
\end{figure}

We can then investigate the systematic behavior of $\ccbar$ pair
distributions as a function of $\kts$.  Fig.~\ref{ccbarpairplusktdist}
shows the transverse momentum distributions of all pairs and also
diagonal pairs. The symbol legends are the same as that in 
Fig.~\ref{cquarkplusktdist} and Fig.~\ref{cquarkplusktrapiditydist}.
The distributions for all pairs (dominated
by the off-diagonal pairs) shown by the corresponding solid
curves are
significantly broader than those for the diagonal pairs at $\kts$ = 0
because the pair $p_T$ = 0 for the LO events.  However, that difference
decreases as $\kts$ increases, because the quarks 
in the off-diagonal pairs receive
uncorrelated $k_T$-kicks.  This is verified by a linear fit to the
$\pts$ of the pair distributions. For the full range of generated
diagonal pairs, one obtains 

\begin{equation}
\pts_{diagonal \; \ccbar } \; = 1.2 \; \gev^2 + 4 \kts,
\label{diagonalccbarptsquaredvskt}
\end{equation}
to be compared with
\begin{equation}
\pts_{all \; \ccbar} \; =  4.9 \; \gev^2  + 2 \kts
\label{allccbarptsquaredvskt}
\end{equation}
for the sample with all pairs.

It is also evident that the diagonal pair sample exhibits some scatter
at the larger $p_T$ values due to statistical fluctuations limited by
the total number of generated events which is of order $10^4$.  This
limitation does not appear until much higher $p_T$ for the all pairs
sample with total number of order $10^8$.  The effect is evident if
we truncate the high-$p_T$ events in the diagonal pair sample.  The
resulting width satisfies

\begin{equation}
\pts_{diagonal \; \ccbar }(p_t < 5 \gev)  \; \approx 0.8 \; \gev^2 + 4 \kts.
\label{diagonalccbarptsquaredvsktlowpt}
\end{equation}
This parameterization will be useful in comparison with initial PHENIX
data in pp and d-Au interactions, where the measured $p_T$ of $\J$ are limited
to the $p_T$ range in Eq. \ref{diagonalccbarptsquaredvsktlowpt}.

\begin{figure}[hbt]
\vskip 1.0truecm
\epsfig{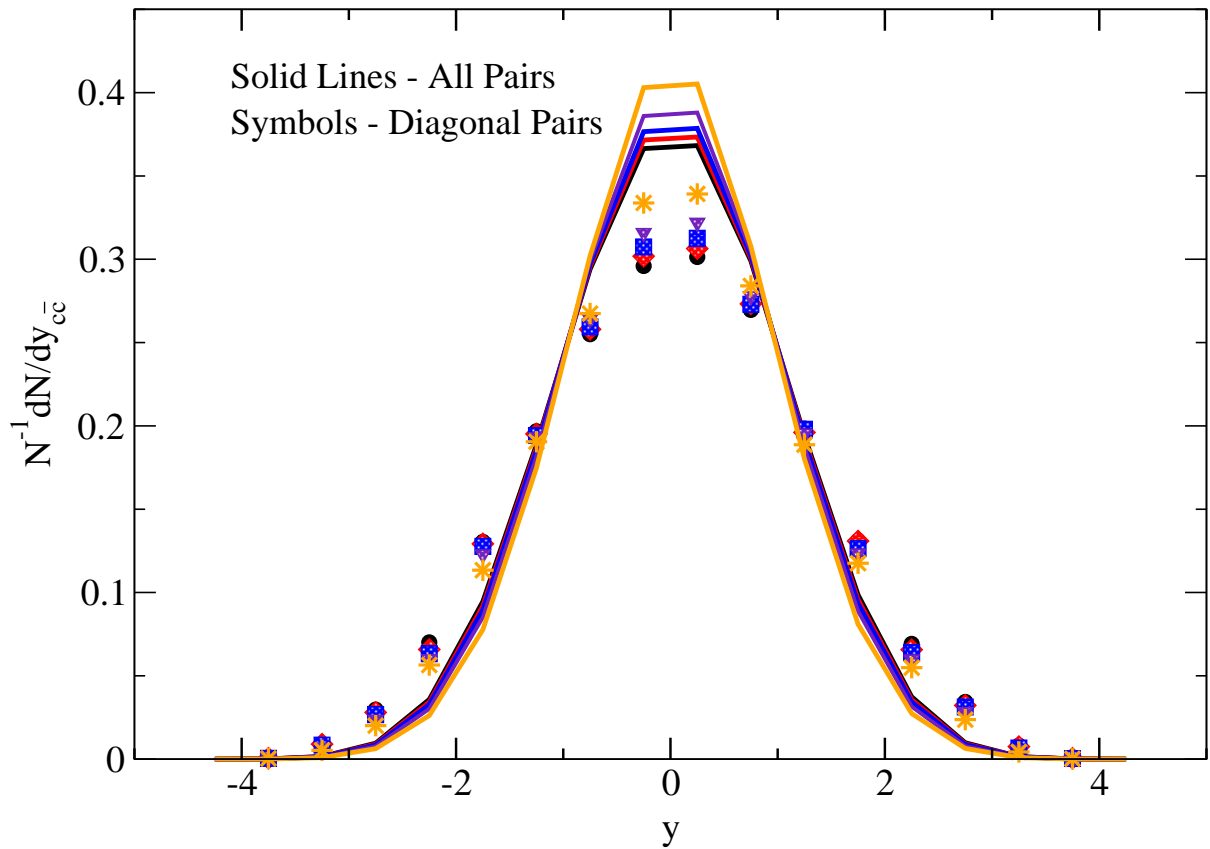}
\caption{(Color online) $\ccbar$ pair rapidity distributions including
intrinsic
$k_T$.  Shown are all pairs (solid lines) and diagonal pairs (symbols,
as in Fig.~\ref{cquarkplusktrapiditydist}).
}
\label{ccbarpairplusktrapiditydist}
\end{figure}

The rapidity spectra of the $\ccbar$ pairs are shown in 
Fig.~\ref{ccbarpairplusktrapiditydist}, again using the same symbol
legends to indicate $\kts$.  One sees that
there is not a dramatic effect in the range of $\kts$ up to 2.0~$\gev^2$.
As $\kts$ increases, the width $\Delta y$ decreases by less than
10 \% for both all pairs and diagonal pairs.


\subsection{Thermal plus flow charm quark distributions}
\label{thermalquark}
We also have generated a set of charm quark momenta to investigate the
opposite scenario, in which the interaction with the partonic constituents
in a deconfined region is sufficiently strong to result in their thermalization
in the expanding medium.  This scenario is not favored by conventional
arguments based on small heavy-quark cross sections and the inhibition of
medium-induced radiation due to the dead-cone effect \cite{Dokshitzer:2001zm}.
Analysis of high $p_T$ leptons in Au-Au collisions as a measurement of
charm quark production are consistent with spectra unmodified from
the initial pQCD calculations \cite{Adcox:2002cg}.  However, it was 
subsequently shown \cite{Batsouli:2002qf} that the resulting spectra
of heavy-quark mesons and electrons for $p_T < 3 ~GeV$ 
cannot distinguish between the two
different scenarios. Recently the possibility of heavy-quark thermalization
was found to be accelerated by interactions involving resonant hadronic
states in a plasma \cite{vanHees:2004gq}.  In any event, it is interesting
to consider such a heavy-quark momentum spectrum in the kinetic formation
process to illustrate the impact of this very different scenario 
on the resulting bound state
spectra.

To generate this set of charm quark momenta, we use the distribution
which follows from numerical results of transverse-boosted Bjorken boost
invariant hydrodynamics \cite{Schnedermann:1993ws},

\begin{equation}
\label{blastwave}
\frac{dN}{d{p_T}^2} \propto m_T \int_0^R{r dr I_0 \left( \frac
{p_T \sinh{y_T(r)}}{T}\right) K_1 \left(\frac{m_T \cosh {y_T(r)}}{T} \right)},
\end{equation}
where the transverse mass $m_T \equiv \sqrt{m^2+{p_T}^2}$
and $R$ is the transverse system size. 
The hadronization temperature $T$ and the rapidity of transverse expansion
$y_T(r)$ are determined from a fit to the light hadrons.  Ref. 
\cite{Batsouli:2002qf} determined $T = 128\;$MeV and $y_T(r) = 
\tanh^{-1}(\beta_T(r))$ with a linear boost profile for the transverse
expansion velocity, $\beta_T(r) = \beta_{max} (\frac{r}{R})$ with
$\beta_{max} = 0.65$ for central collisions.     
This form has also been used to fit the transverse momentum spectrum
of hadrons in a statistical model in which the freeze-out and hadronization
temperatures are identical \cite{Bugaev:2002zk}. In this case the linear profile
of the transverse expansion was taken to vary with transverse rapidity
rather than velocity, but for typical $\beta_{max}$ values the difference
is not significant.  For the calculations here, we have used the 
linear profile in transverse rapidity.
We also consider an alternate set of parameters, 
$T = 170$~MeV and $\beta_{max}$ = 0.5 which emerge from a coalescence
study of thermal partons which also fit the light hadron spectra at
RHIC \cite{Greco:2003mm}.

\begin{figure}[hbt]
\vskip 1.0truecm
\epsfig{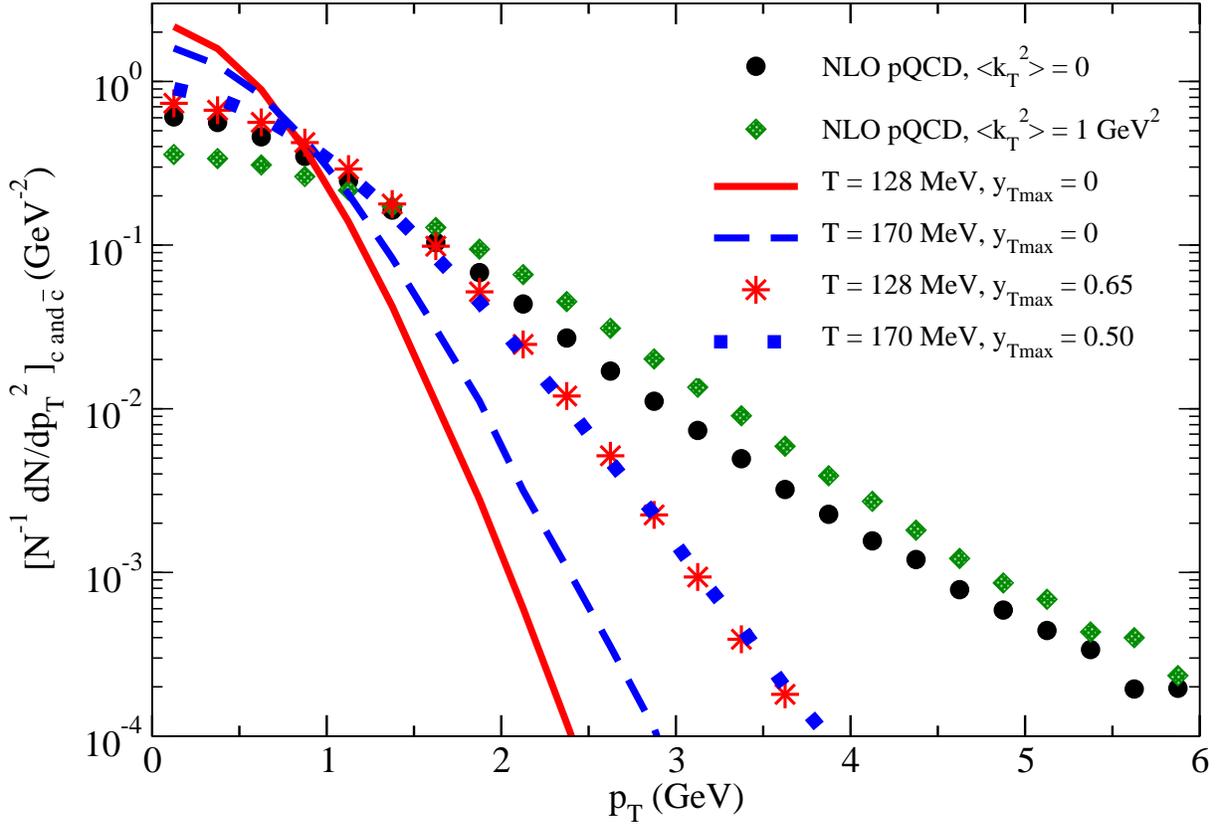}
\caption{(Color online) Thermal plus flow charm quark transverse momentum distributions.
Shown are thermal without flow (solid and dashed lines), thermal with
flow (star and dotted line), and pQCD (symbols).
}
\label{thermalcharm}
\end{figure}

Fig.~\ref{thermalcharm} shows the transverse momentum spectra of
charm quarks which result from Eq. \ref{blastwave} using the parameter
sets $(T, {y_T}_{max})$ = (128 MeV, 0), (170 MeV, 0),   
(128 MeV, 0.65) and (170 MeV, 0.50).  For no flow, the distributions 
for different temperatures are of course
different.  However when augmented by the corresponding transverse flow 
the resulting
distributions are almost identical, since both ${y_T}_{max}$ values have
been adjusted to fit the observed flow at the corresponding freezeout
temperatures.  

The pQCD distributions shown for
comparison are much broader and have 
an obvious non-thermal behavior for large $p_T$.  
The widths can be characterized by the calculated $\pts$ values 
of 0.47, 0.67, 1.35, and 1.24 $\gev^2$ for the thermal and thermal
plus flow distributions, compared with 2.45 and 3.45 $\gev^2$ for
the pQCD examples.

\begin{figure}[hbt]
\vskip 1.0truecm
\epsfig{file=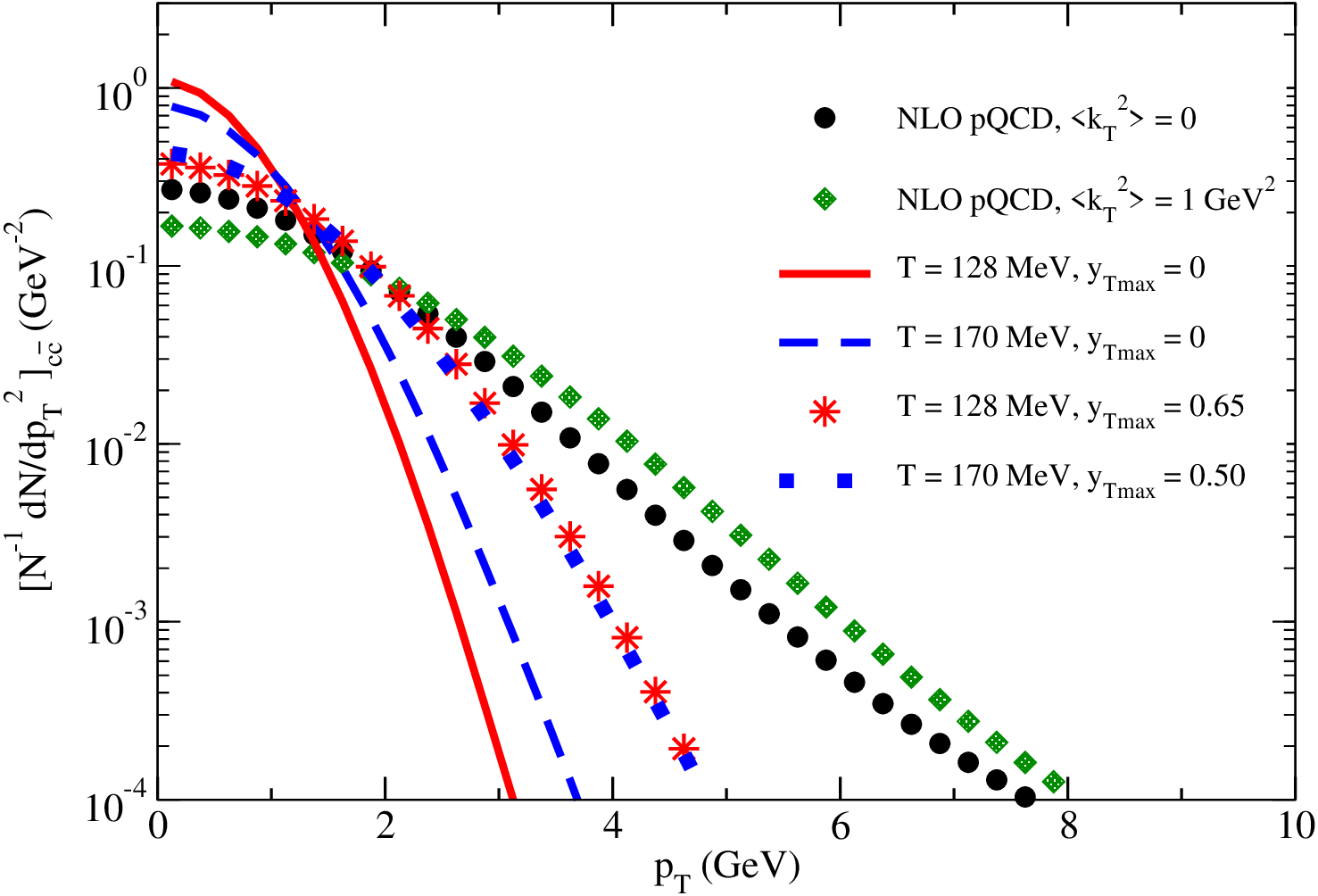, width=16cm}
\caption{(Color online) Thermal plus flow pair transverse momentum distributions.
Shown are thermal without flow (solid and dashed lines), thermal with
flow (star and dotted line), and pQCD (symbols).
}
\label{thermalccbarpairspt}
\end{figure}

The transverse momentum spectra of the thermal plus flow $\ccbar$ pairs
are shown in Fig.~\ref{thermalccbarpairspt}. In this case there is no
set of diagonal pairs since all identity has been lost in the thermalization
process.  The pair distribution widths are twice those of the corresponding
distributions for single charm quarks, as expected due to the 
uncorrelated relative azimuthal angles.  Their $\pts$ values 
are 0.93, 1.32, 2.70, and 2.46 GeV$^2$, respectively.  Again these are
much smaller than the pQCD comparison distributions, which have
$\pts$ of 4.9 and 6.9 GeV$^2$.

\section{Details of kinetic formation model calculations}
\label{kineticpredictions}

We generate the transverse momentum and rapidity spectra of the formed
$\J$ according to Eq. \ref{formdist}.  For each $\ccbar$ pair, we transform to
the pair CM-system and weight the formation event by ${\it v_{rel}}~\sigF$.
The formation cross section $\sigF$ is obtained via detailed balance
from the dissociation cross section $\sigD$, 

\begin{equation}
\sigF = \frac{48}{36} \sigD \frac{\left(s-{M_{\J}}^2\right)^2}
{s\left(s-4 {m_c}^2\right)},
\end{equation}
in terms of the charm quark mass $m_c$ and the energy invariant 
$s$.  $\sigD$ is taken from Eq. \ref{eqsigma}, with modified threshold 
gluon momentum to account for the finite mass of the $\J$.  

The angular distribution of the $\J$ in the $\ccbar$ rest frame is 
 taken from the corresponding behavior of atomic
or nuclear \cite{Wong:2001kn} photodissociation which gives
\begin{equation}
\frac{d\sigF}{d^3p} \propto {\sin^2{\theta}}.
\label{angulardist}
\end{equation}
The sensitivity to the high energy behavior of $\sigF$ is probed by
imposing a maximum energy cutoff parameter $d_{max}$ 
\footnote
{This is equivalent to the cutoff imposed by
the color evaporation model at the open-flavor threshold.}
which limits the
invariant mass of the $\ccbar$ pairs which can initiate the formation
of a $\J$:
\begin{equation}
\sqrt{s_{\ccbar}} < M_{\J} +d_{max}
\label{dmax}
\end{equation}

\begin{figure}[hbt]
\vskip 1.0truecm
\epsfig{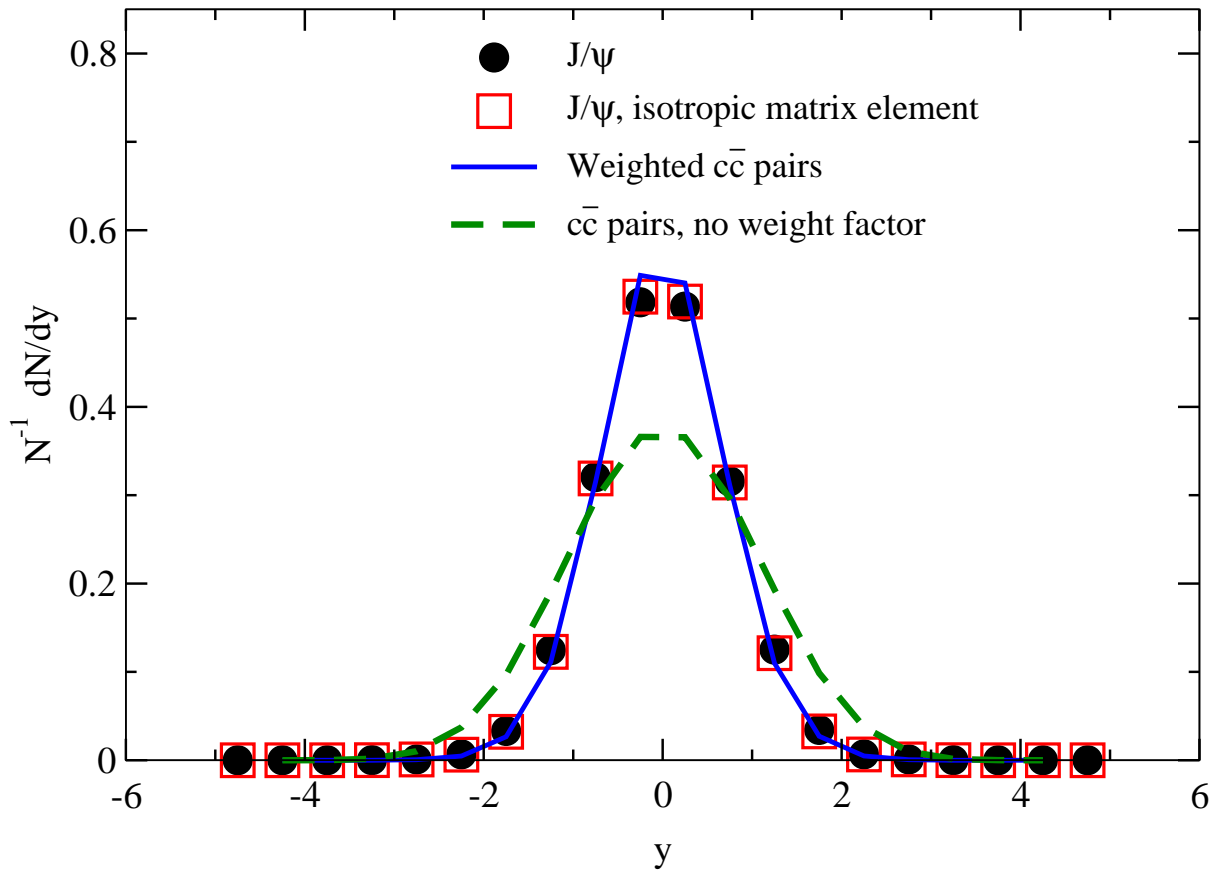}
\caption{(Color online) Comparison of rapidity spectra for $\ccbar$ pairs and formed
$\J$, all using default parameters.}
\label{rhicyspectracomparison}
\end{figure}

\begin{figure}[hbt]
\vskip 1.0truecm
\epsfig{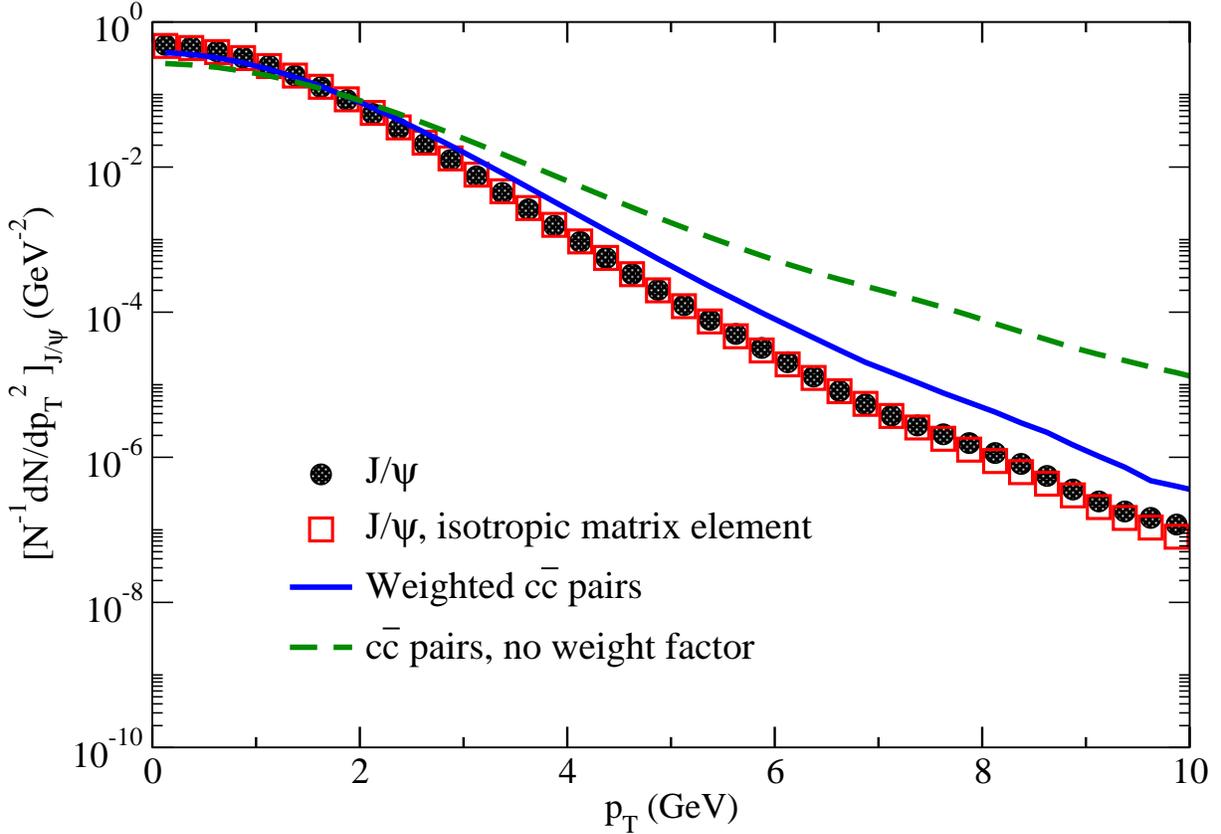}
\caption{(Color online) Comparison of transverse momentum spectra for $\ccbar$ pairs and
formed $\J$, all using default parameters.}
\label{rhicptspectracomparison}
\end{figure}

The resulting rapidity and transverse momentum spectra are given by
the black circles compared in
Fig.~\ref{rhicyspectracomparison} and
Fig.~\ref{rhicptspectracomparison}, respectively. 

Our choice of initial default parameters set the binding energy 
$\epsz$ = 0.63 GeV, the mass cutoff parameter
$d_{max}$ = 3.0 GeV, and no $k_T$ kick ($\kts$ = 0). The curves refer to
a sample of $10^4$ independent $\ccbar$ pairs, resulting in $10^8$
possible pairings.

 The impact  on these distributions of the various  dynamical
assumptions is shown by three additional curves:
\begin{itemize}
\item $\J$, isotropic matrix element (empty squares): the angular
  dependence in Eq.~\ref{angulardist} is taken to be flat;
\item weighted $\ccbar$ pairs (solid lines): we plot the kinematical
  variables of the $\ccbar$ pairs weighted by the $\J$ formation 
probability ${\it v_{rel}}~\sigF$, which differs from the formation
process by elimination of exact kinematics relating a $\ccbar$ pair
to the spectrum of $\J$ which follow from the in-medium formation
reaction;
\item  $\ccbar$ pairs (dashed lines): we plot the kinematical
  variables of the $\ccbar$ pairs which result from all combinations
of charm quarks as produced in the pQCD processes.
\end{itemize}
One sees that for both spectra the $\J$ formation is not sensitive to the 
angular dependence of the amplitude, as evidenced by the equality of the
circle and square plots (except a small effect at the largest values
of transverse momentum).  The primary effect of the formation dynamics
is to produce a narrowing of all spectra, compared with that of the
 $\ccbar$ pairs (dashed line).  For the rapidity spectra, 
even the weighted
$\ccbar$ pairs (solid line) are identical to the $\J$.  

\begin{figure}[hbt]
\vskip 1.0truecm
\epsfig{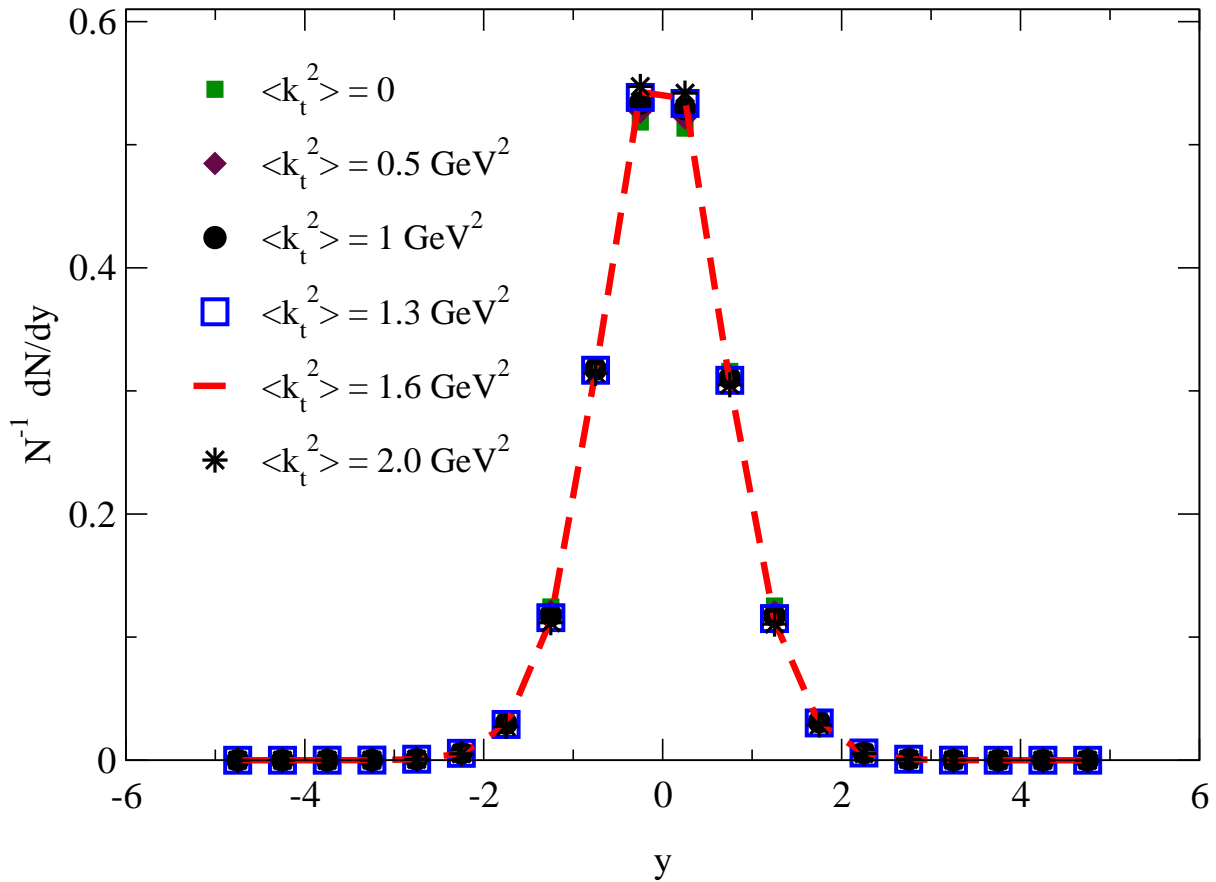}
\caption{(Color online) Effects of intrinsic $\kts$ on the $\J$ formation rapidity
spectrum.}
\label{rhicyspectravskt}
\end{figure}

\begin{figure}[hbt]
\vskip 1.0truecm
\epsfig{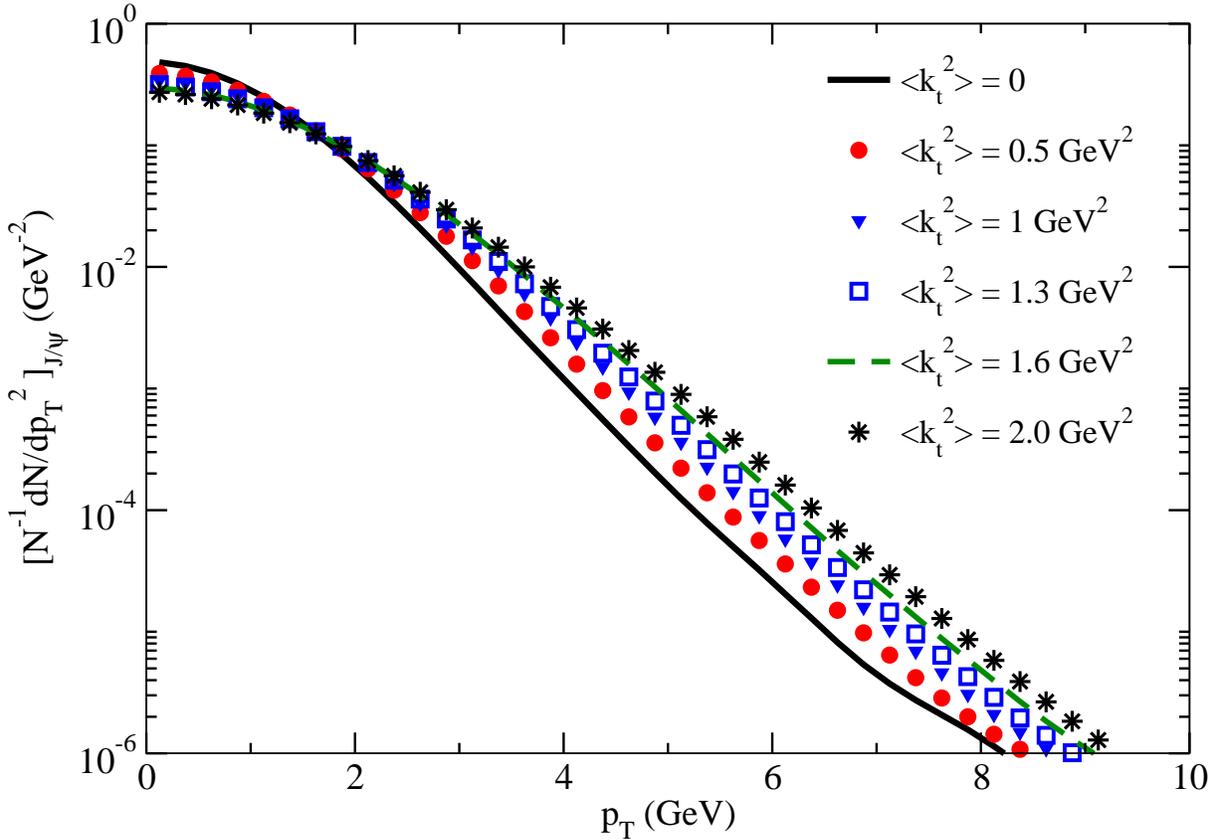}
\caption{(Color online) Effects of intrinsic $\kts$ on the $\J$ formation transverse
momentum spectrum.}
\label{rhicjpsiptspectravskt}
\end{figure}

The effects of intrinsic $\kts$ on the $\J$ spectra are shown in 
Fig.~\ref{rhicyspectravskt} and Fig.~\ref{rhicjpsiptspectravskt}.

The rapidity spectra are essentially independent of $\kts$.  The transverse
momentum spectra increase with $\kts$, as expected.  We can parameterize
this effect in terms of $\pts$ for the formed $\J$, as 

\begin{equation}
\pts_{\J} \; = 2.4 \; \gev^2 + \kts.
\label{jpsiptsquaredvskt}
\end{equation}

It is interesting to compare this form with the corresponding
behavior of all $\ccbar$ pairs in Eq. \ref{allccbarptsquaredvskt},
in which the value at $\kts$ = 0 is larger (4.9 $\gev^2$) as is the
rate of increase (2).  A fit of an intermediate case, using
all $\ccbar$ pairs weighted with the formation cross section 
${\it v_{rel}}~\sigF$ yields
an intermediate result, 

\begin{equation}
\pts_{ccbar(weighted)} \; = 3.2 \; \gev^2 + 1.5 \; \kts.
\label{ccbarformationbiasedptsquaredvskt}
\end{equation}

\begin{figure}[hbt]
\vskip 1.0truecm
\epsfig{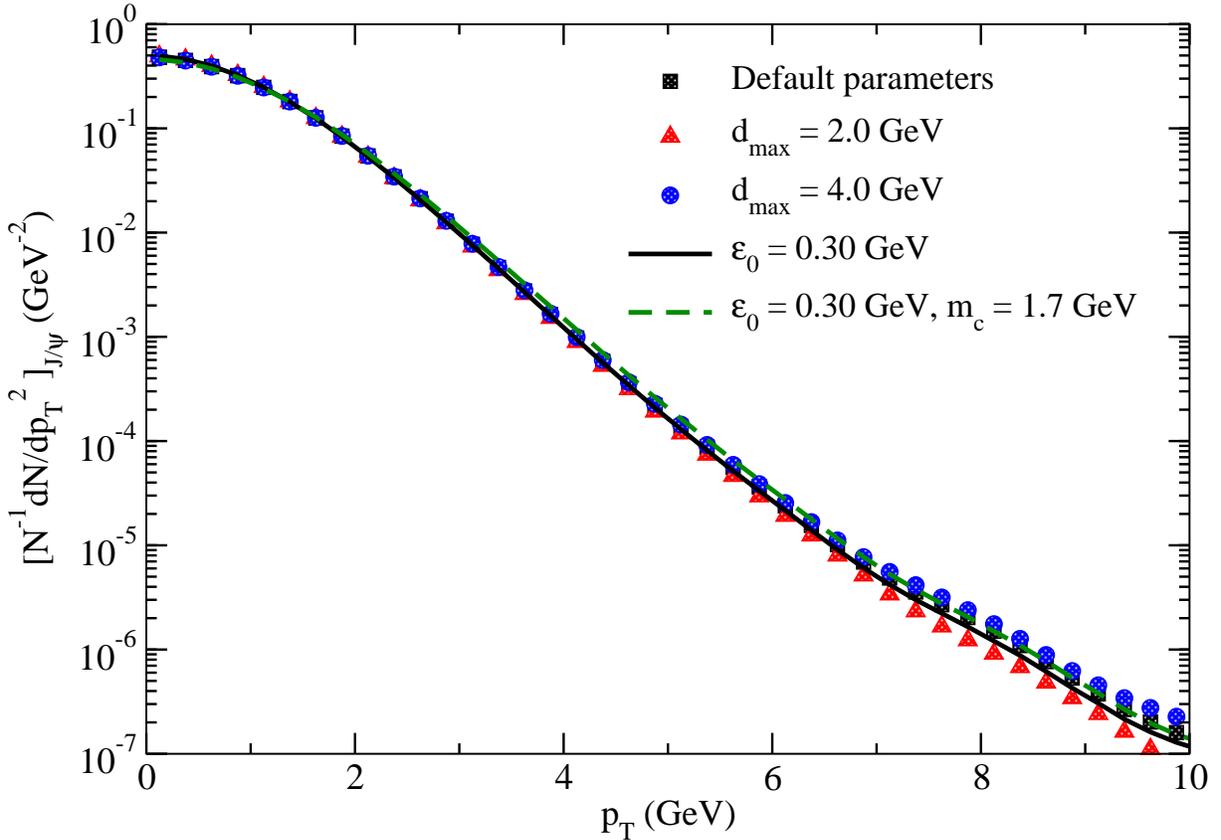}
\caption{(Color online) Effects of parameter variations on the $\J$ formation transverse
momentum spectrum.}
\label{jpsiptparametervariation}
\end{figure}

\begin{figure}[hbt]
\vskip 1.0truecm
\epsfig{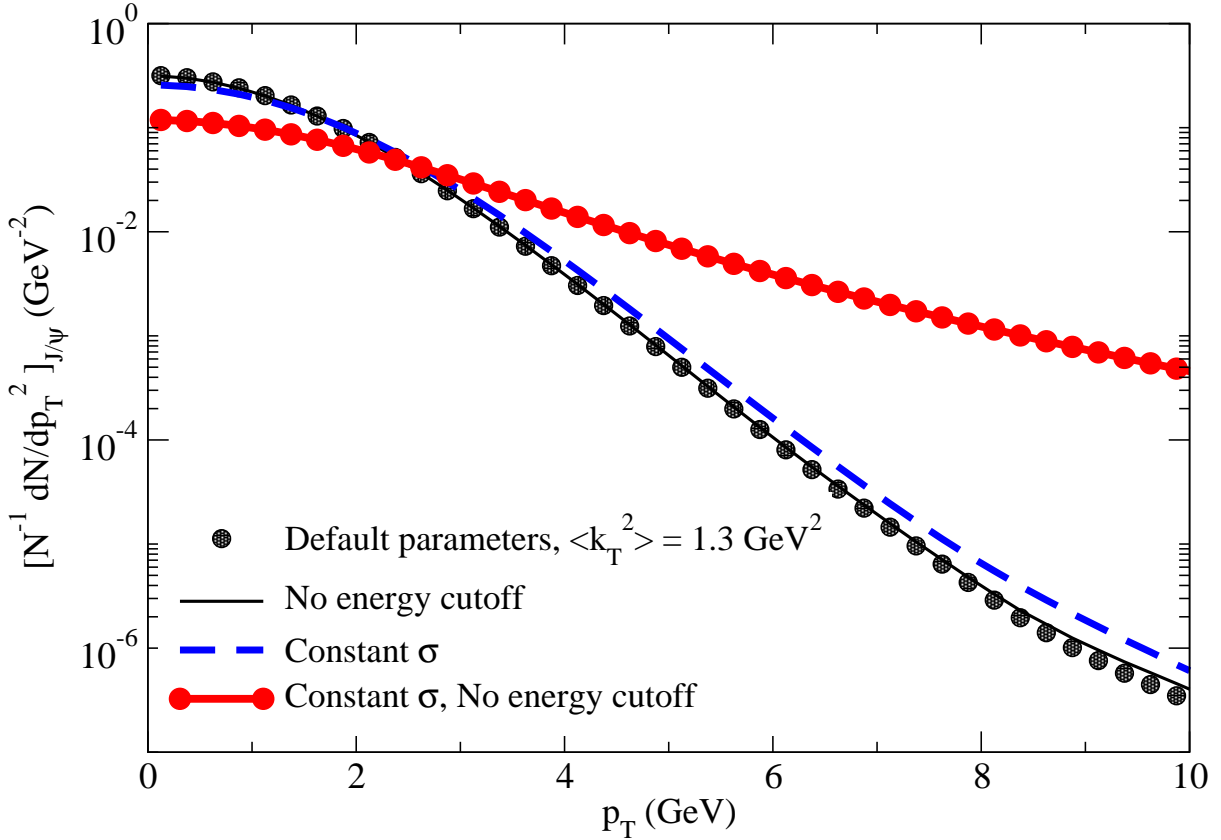}
\caption{(Color online) Dependence of the $\J$ formation transverse
momentum spectrum on the formation cross section.}
\label{jpsiptconstsigma}
\end{figure}

Next we show the sensitivity of the formation results to variations in
some of the default parameters.  Fig.~\ref{jpsiptparametervariation} 
shows the results
for the normalized transverse momentum spectra.  
The first three
curves show variation in $d_{max}$ (symbols).  The final two curves
 (solid and dashed lines) 
show the effects of reducing the binding energy to $\epsz$ = 300 MeV, 
more than a factor of two below the vacuum value.  In addition, the
last curve has the quark mass increased to $m_c$ = 1.7 GeV, which
restores the $\J$ mass to its vacuum value.  One sees that overall these
changes produce negligible variations in the $p_T$ spectra.
Lastly, we look at extreme variations in the functional form of the cross
sections to determine the degree of stability of the $\J$ formation 
$p_T$ spectra.  Fig.~\ref{jpsiptconstsigma} shows some of these effects.
The effect of omitting an energy cutoff ($d_{max} \rightarrow \infty$)
is shown by the thin solid line,
which is essentially unchanged from the default parameter result
(shaded circles) with
$d_{max}$ = 3 GeV.  This might be understood from the behavior of
the OPE-cross sections, since the entire region around the peak values is
generally contained within the default parameter choices. We next consider
the effect of a cross section independent of $\sqrt{s}$, 
shown by the dashed line. (The
calculation is virtually unchanged whether one chooses $\sigF$ or
$\sigD$ to be constant).  This choice of cross section produces
only a small change in the $\J$ transverse momentum spectrum.  The most
radical departure is to use constant cross sections and no energy cutoff.
These results are shown by the solid line and circle, and do produce
a significant change in the $p_T$ spectrum. We would certainly not
expect, however, that the $\J$ formation probability would remain
independent of the $\ccbar$ pair relative momentum throughout the entire 
range populated by the pQCD initial production process.  To the contrary,
one would expect that the large invariant-mass pairs would preferentially
hadronize into the higher-mass states in the open charm spectrum.

\begin{figure}[hbt]
\vskip 1.0truecm
\epsfig{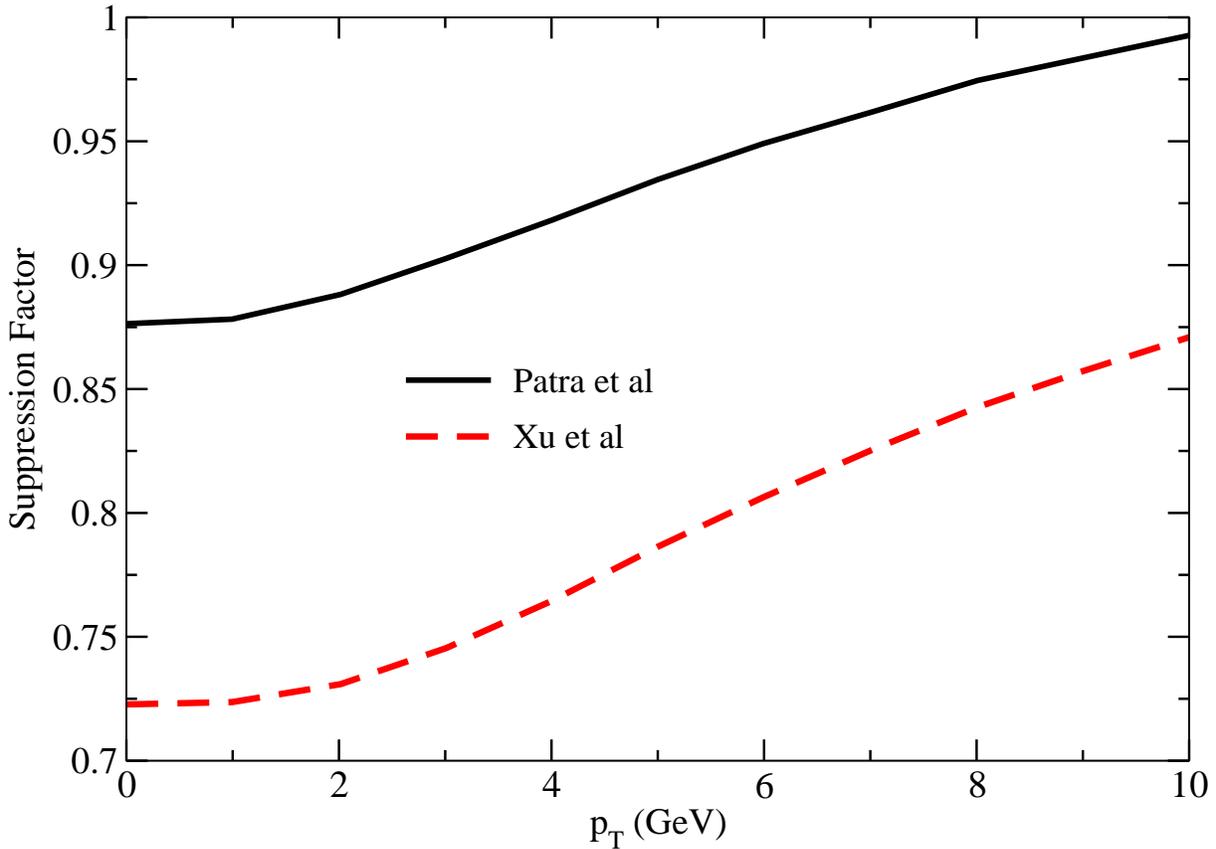}
\caption{(Color online) In-medium $\J$ suppression factors.}

\label{rhicjpsisuppfactors}
\end{figure}

\begin{figure}[hbt]
\vskip 1.0truecm
\epsfig{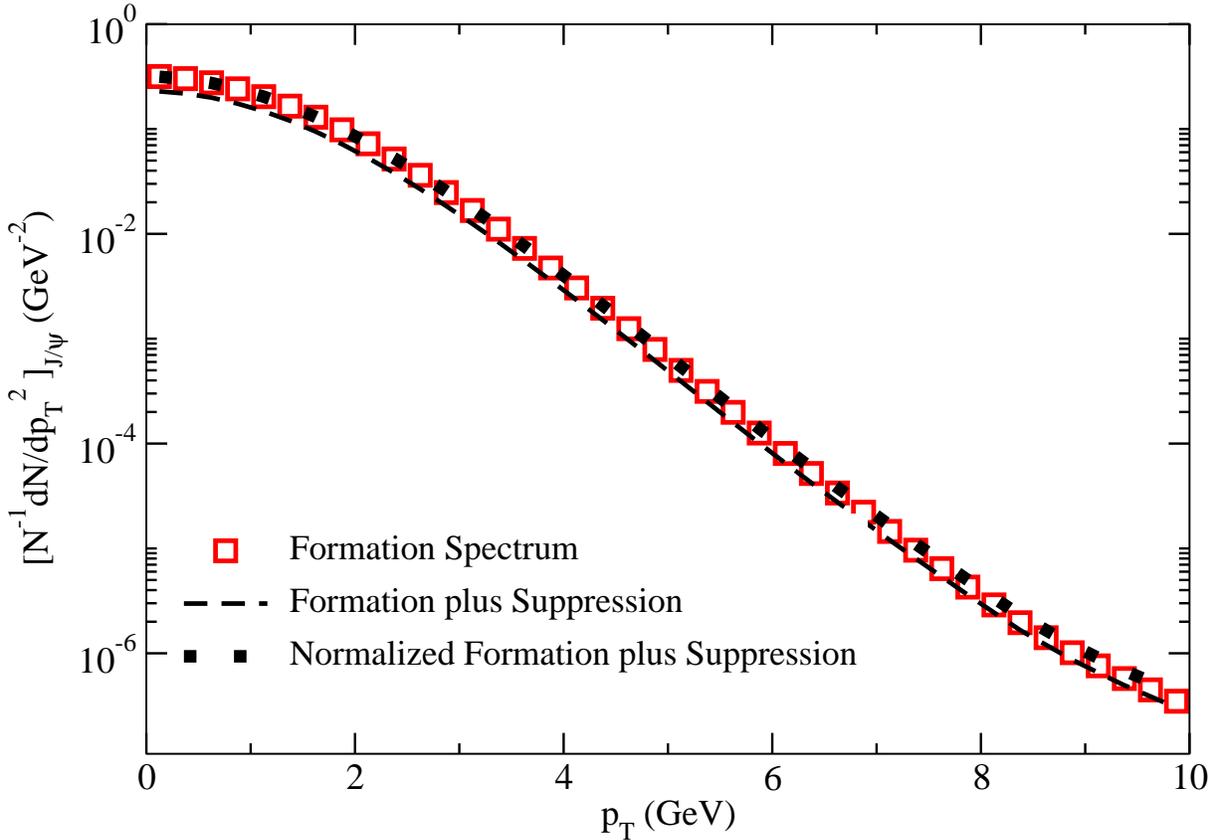}
\caption{(Color online) The in-medium $\J$ normalized formation plus suppression spectrum
shown by the small filled squares is compared with the normalized 
formation-only spectrum shown by the large open squares.
}
\label{rhicjpsisuppnorm}
\end{figure}

Finally, we investigate the effects of in-medium dissociation 
on the formed $\J$ spectrum.  This dissociation will
influence the final $p_T$ spectrum in two competing scenarios.  In the
first, one expects that high-$p_T$ $\J$ which are initially produced
will originate preferentially in central collisions, where the Cronin 
effect is most efficient, and thus be subject to a large suppression.  
However, one also expects that the $\J$ with large
$p_T$ will preferentially escape the deconfinement region. In a recent
study \cite{Zhu:2004nw} incorporating these effects for $\J$ propagating 
in a region
of deconfinement at both SPS and RHIC energy, it was found that
this second ``leakage" effect dominates.  We will use numerical
values taken from 
initial studies of this effect \cite{Xu:1995eb} and updated versions 
\cite{Patra:2004wf, 
Patra:2005yg} in the form of $p_T$-dependent suppression factors 
for $\J$ initially present in an equilibrating and expanding
parton gas.
For the application to our formation spectrum,
we need a  suppression factor for $\J$ formed
at a continuum of initial times.  However, the time of formation
is biased toward early times when the charm quark density is
maximum.  Hence we use two typical suppression factors from
Ref. \cite{Patra:2004wf}, which have their minimum value at $p_T$ = 0 
and approach unity for $p_T \approx 10~\gev$.
These are shown in Fig.~\ref{rhicjpsisuppfactors}. 
The effect of these suppression factors on the formation spectra is
shown in Fig.~\ref{rhicjpsisuppnorm}. One sees that the normalized 
formation spectra are suppressed in the low-$p_T$ region by amounts
from 10 - 25\%.  However, the {\it normalized} spectrum including both 
formation and suppression is virtually unchanged in the low-$p_T$ 
region (which dominates the normalization factor), and the entire
effect of dissociation is pushed out to very large $p_T$.  Hence the
normalized formation spectra alone provide sufficiently accurate
predictions, at least in the region $0 < p_T <~6~GeV$ where the
majority of $\J$ are found.  This effect can be quantified by the
small increase in $\pts$, from 3.59 $GeV^2$ to 3.69 $GeV^2$.

\section{Predictions for Au-Au interactions at RHIC}
\subsection{Charm quark distributions from pQCD}
\label{pqcd}
\begin{figure}[hbt]
\vskip 1.0truecm
\epsfig{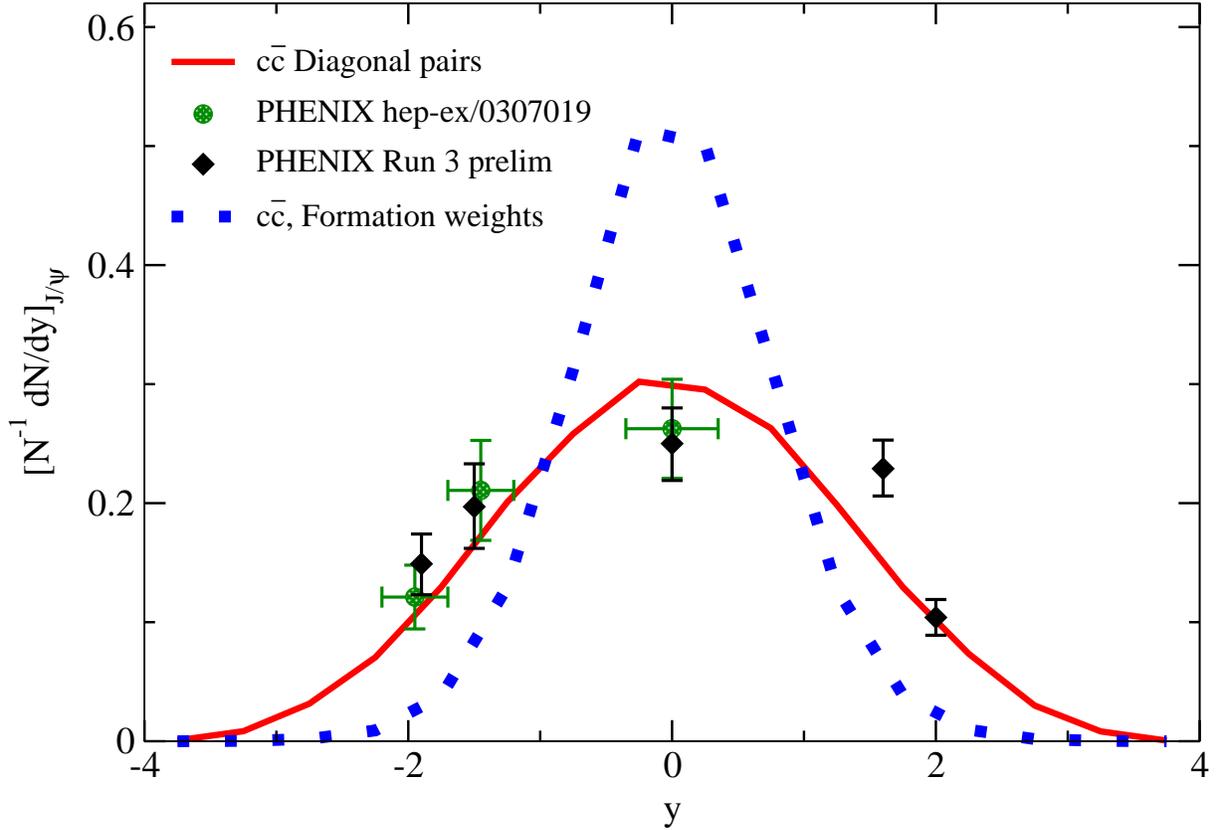}
\caption{(Color online) Rapidity spectra of $\J$ production in pp interactions
at 200 GeV.
}
\label{jpsippyspectraplusdata}
\end{figure}

In order to make predictions of the kinetic formation model for 
$\J$ at RHIC, we need to fix the $\kts$ parameter.  We start with
$\J$ production in pp interactions at 200 GeV.  There are PHENIX data on the
rapidity and transverse momentum spectra from both run2 \cite{Adler:2003qs} and
run3 \cite{deCassagnac:2004kb}, which we exhibit in Fig.
\ref{jpsippyspectraplusdata}. 
Since multiple $\ccbar$ pair production in
pp interactions is negligible, we consider only the diagonal $\ccbar$ 
pairs. These are shown by the solid curve, and are seen to describe 
the data reasonably well. (Recall from Fig.~\ref{ccbarpairplusktrapiditydist} 
that the
$\ccbar$ pair rapidity spectra are essentially independent of $\kts$.)
One can interpret this agreement as 
consistency with any mechanism in which $\J$ is produced via hadronization
of initially-produced $\ccbar$ pairs.  Also shown by the dotted curve 
are the pairs modified by the in-medium formation probability, proportional
to $\sigF$.  This process of course cannot occur in a pp interaction, and 
it is gratifying to see that it is incompatible with the data.

\begin{figure}[hbt]
\vskip 2.0truecm
\epsfig{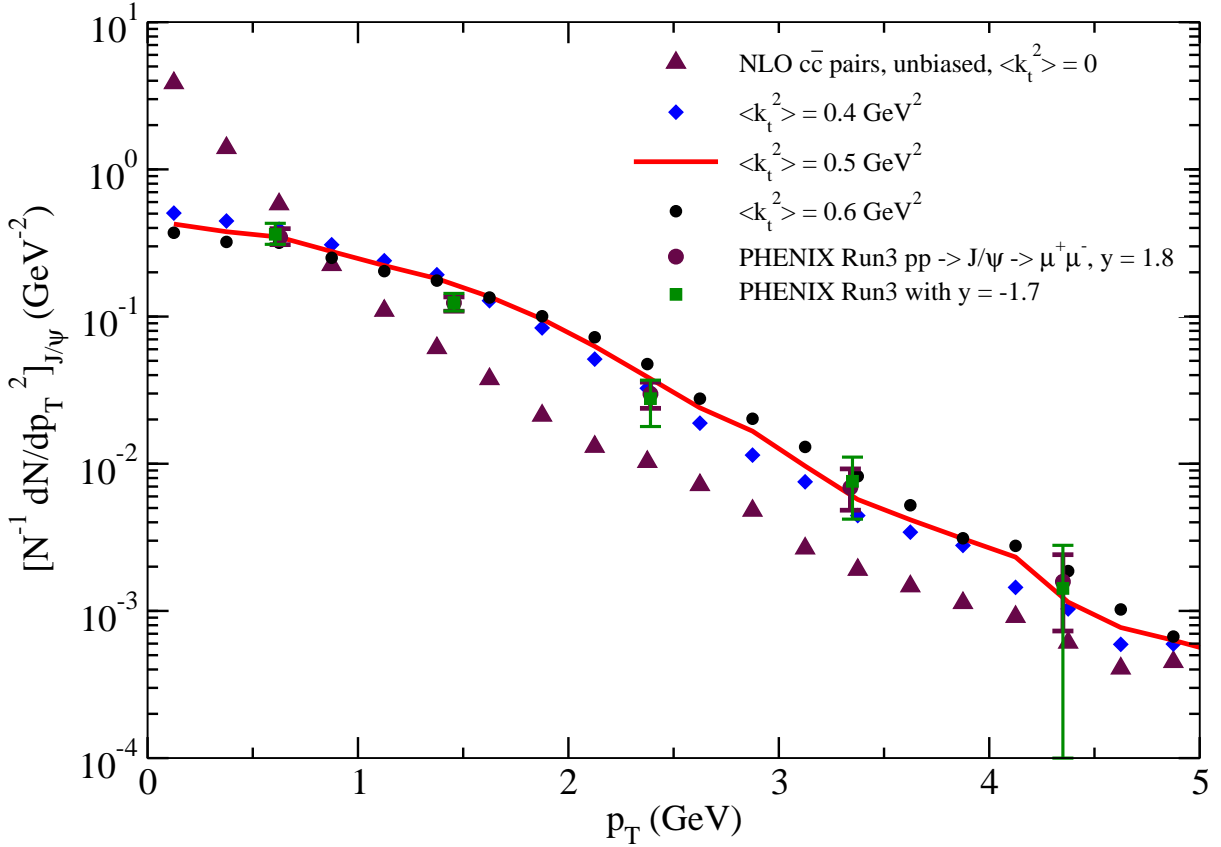}
\caption{(Color online) Transverse momentum spectra of $\J$ production in pp interactions
at 200 GeV.
}
\label{jpsippptspectraplusdata}
\end{figure}

The transverse momentum spectrum is shown in Fig.~\ref{jpsippptspectraplusdata}.
The PHENIX dimuon data are presented for both positive and negative 
rapidity, which have smaller uncertainties than the central dielectron 
data.  The solid triangles are the pQCD $\ccbar$ diagonal pairs, which
of course must be augmented by some\;$\kts$ to include initial state
and confinement effects.  The solid line and adjacent circles and diamonds 
represent $\kts\; = 0.5 \pm 0.1\; \gev^2$, and provide an acceptable fit
to the data (fluctuations at large $p_T$ are statistics limited from
the number of generated diagonal pQCD events).

There are also preliminary data on $\J$ production in d-Au interactions 
\cite{deCassagnac:2004kb}. The width of the $p_T$ spectra is found to
be larger than that measured in pp interactions.

\begin{equation}
\pts_{d-Au} - \pts_{pp}\; =\; 
\begin{cases}
1.77 \pm 0.35 \;\gev^2 \;\;(y = -1.7)& \\ 
1.29 \pm 0.35 \;\gev^2 \;\;(y = +1.8)&
\end{cases}
\label{dAupt}
\end{equation} 

This broadening of $p_T$ distributions for particles produced on nuclear
targets is well known, and fits a natural pattern which emerges from
initial state elastic scattering of a projectile in the nuclear target 
\cite{Gavin:1988tw}. For p-A (or d-A) interactions, the increase can be
expressed as a change in $\pts$, as

\begin{equation}
\pts_{pA} - \pts_{pp}\; = \lambda^2\; [\bar{n}_A - 1], 
\label{pApt}
\end{equation}
where $\bar{n}_A$ is the impact-averaged number of inelastic interactions
of the projectile in nucleus A, and $\lambda^2$ is the square of the
transverse momentum transfer per collision. For a nucleus-nucleus collision,
the corresponding relation is 
\begin{equation}
\pts_{AB} - \pts_{pp}\; = \lambda^2 \;[\bar{n}_A + \bar{n}_B- 2]. 
\label{AApt}
\end{equation}
We use the measured $\J$ broadening in d-Au to determine the appropriate
$\kts$ value for the $p_T$ distribution of initially-produced diagonal
$\ccbar$ pairs through Eq. \ref{diagonalccbarptsquaredvskt} 
or \ref{diagonalccbarptsquaredvsktlowpt}, with the result
$\kts_{d-Au} - \kts_{pp}\; = \;0.4 \pm 0.1$.  (Since the 
measured values for d-Au interactions differ between positive and
negative rapidity, we use their average to partially compensate
for the existence of final-state effects.)
Finally, using $\kts_{pp}$ and $\kts_{d-Au}$ extracted from data,
in combination with Eq. \ref{AApt} leads to $\kts_{Au-Au} \;= 1.3 \pm 0.3 
\;\gev^2$.  We use this value to determine the $p_T$ distribution of
initially-produced charm quarks in Au-Au interactions.  It is 
interesting to also look at the equivalent parameters which 
follow from Eq. \ref{dAupt} and Eq. \ref{pApt}.  One finds 
$\bar{n}_A = 5.4$ for minimum bias d-Au interactions at RHIC
energy (using $\sigma_{pp}$ = 42 mb), which leads to 
$\lambda^2 = 0.35 \pm 0.14 GeV^2$.  We note that the relatively large
uncertainty comes entirely from the difference in $p_T$ broadening
in d-Au between positive and negative rapidity. It is interesting to
see the energy dependence by comparing this value with that extracted 
from $\J$ pA data at 
fixed-target energy \cite{Gavin:1988tw}, $\lambda^2 = 0.12 \pm 0.02 GeV^2$.

\begin{figure}[hbt]
\vskip 1.0truecm
\epsfig{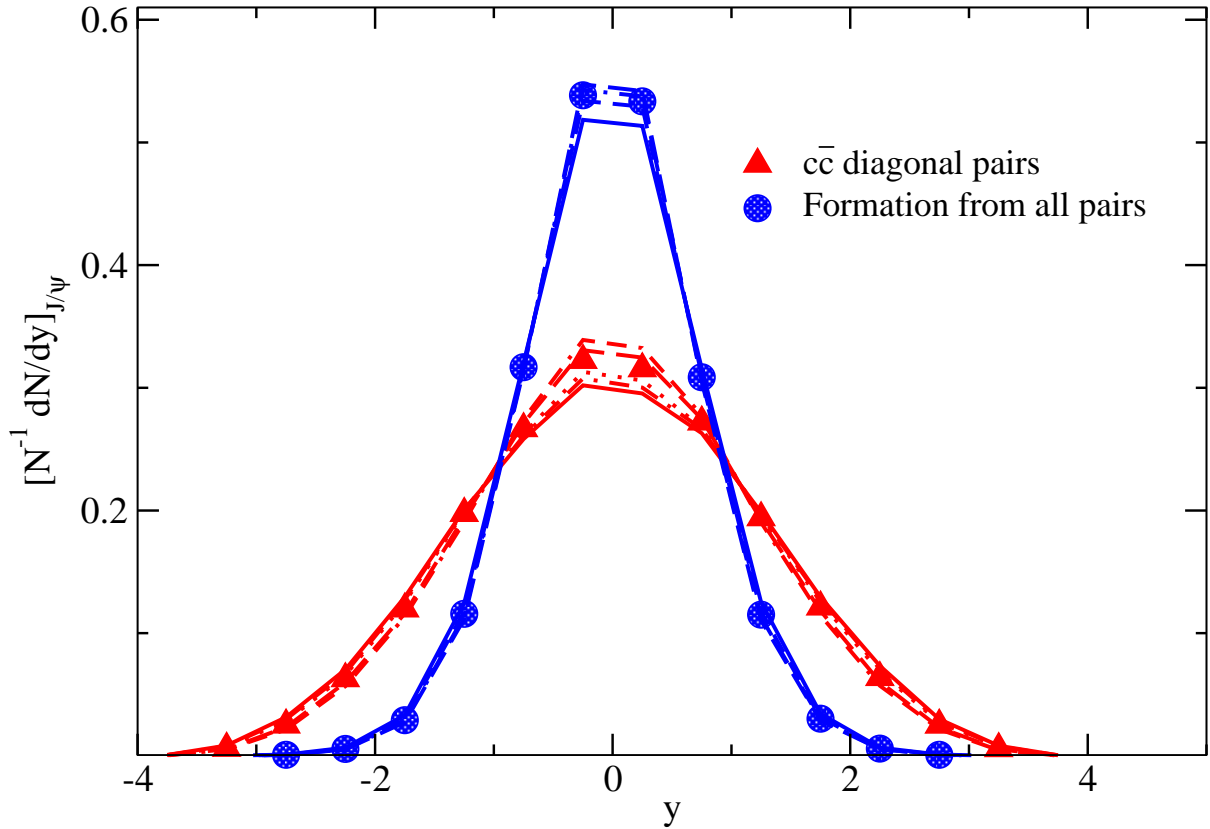}
\caption{(Color online) Predicted rapidity spectra of $\J$ in Au-Au interactions
at 200 GeV. Triangles are for initial production via diagonal
$\ccbar$ pairs.  Circles are for in-medium formation via all
pairs.  Sensitivity of the formation spectrum
to variation of $\kts$ within the range
$\kts = 1.3 \pm 0.3~GeV^2$ is indicated by the spread in the solid lines.
The corresponding spread in the lines for diagonal pairs covers
the range $0~<~\kts~<~2~GeV^2$.}
\label{ypredictions}
\end{figure}

Shown in Fig.~\ref{ypredictions} are the predicted rapidity spectra
of $\J$ in Au-Au interactions at 200 GeV.  
The solid triangles 
use initially-produced diagonal $\ccbar$ pairs with $\kts$ = 1.3 $\gev^2$,
and the various lines indicate the small sensitivity to variations 
of $\kts_{Au-Au}$ within uncertainties.  This distribution is also
virtually the same as those using $\kts_{pp}$, which fit the pp data
as shown in Fig.~\ref{jpsippyspectraplusdata}.  The solid circles and
associated lines show the results of a formation calculation using
all $\ccbar$ pairs which can recombine in the medium.  These spectra 
are substantially narrower, and provide a prediction which signals the
existence of the formation mechanism.  Since the formation process
is largest for central collisions, where dissociation of the initial
production yield should be most efficient, one would expect that
the total $\J$ rapidity spectrum should change from narrow for the
most central collisions to wide for the most peripheral collisions.

\begin{figure}[hbt]
\vskip 1.0truecm
\epsfig{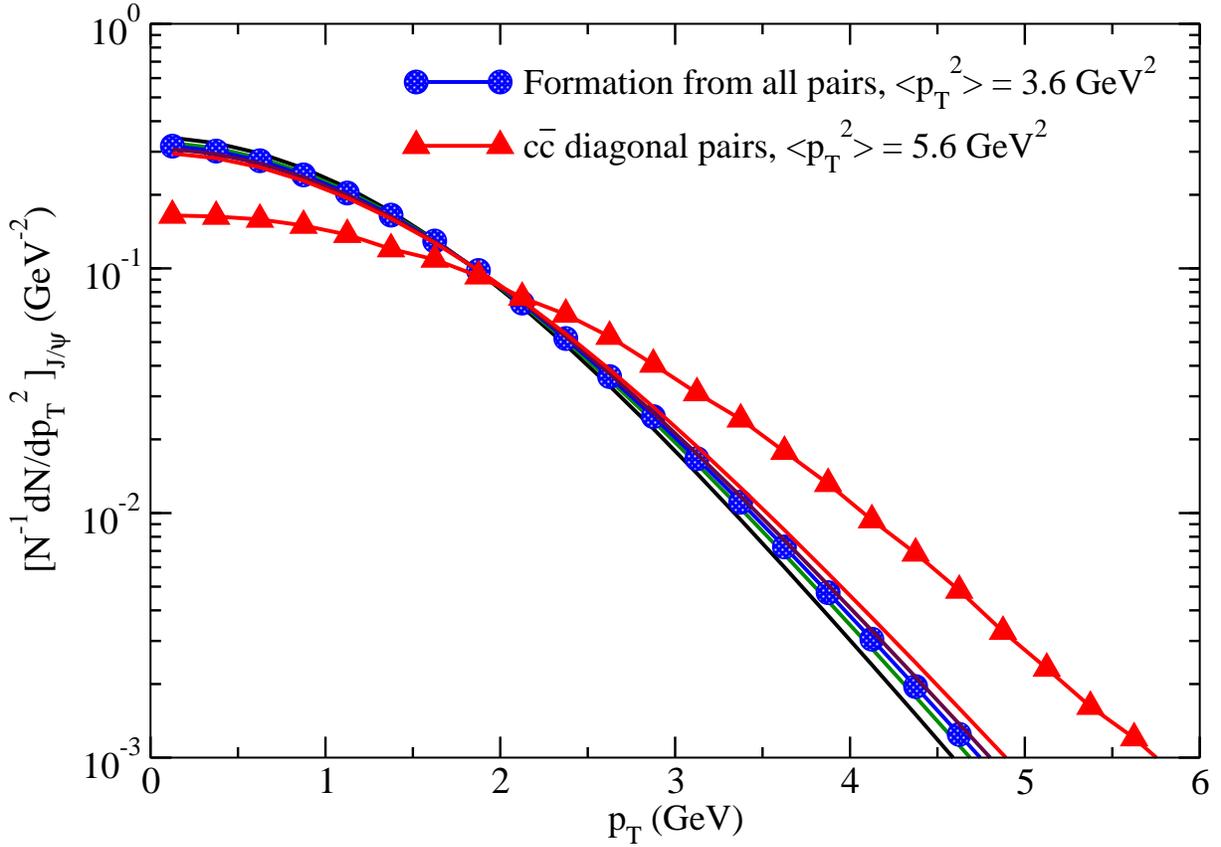}
\caption{(Color online) Predicted transverse momentum spectra of $\J$ in Au-Au interactions
at 200 GeV. Triangles are for initial production via diagonal
$\ccbar$ pairs.  Circles are for in-medium formation via all
pairs.  Sensitivity of the formation spectrum
to variation of $\kts$ within the range
$\kts = 1.3 \pm 0.3~GeV^2$ is indicated by the spread in the solid lines.
}
\label{ptpredictions}
\end{figure}

Fig.~\ref{ptpredictions} shows the corresponding transverse momentum
spectra.  
The solid triangles again 
use initially-produced diagonal $\ccbar$ pairs with $\kts_{Au-Au}$ = 1.3 GeV$^2$.
Note that this distribution is 
wider than those using $\kts_{pp}$, which fit the pp data
shown in Fig.~\ref{jpsippptspectraplusdata}. 
 The solid circles and
associated lines (which include the uncertainty in $\kts_{Au-Au}$) 
show the results of a formation calculation using
all $\ccbar$ pairs which can recombine in the medium.  These spectra
are substantially narrower, and provide another prediction which signals the
existence of the formation mechanism.  The same consideration of centrality
dependence as presented for the rapidity spectra also apply to these
transverse momentum spectra.

\subsection{Charm quark distributions from thermal plus flow}
\label{thermal}

The calculation of $\J$ follows from the same cross section which has been
used for the pQCD charm quark distributions.  We use generated $\ccbar$
events as calculated in  
Sec. \ref{thermalquark}, where it was demonstrated 
 that the charm quark transverse momentum distributions
which follow from thermal plus flow parameters $T = 128$ MeV, $y_{Tmax} = 0.65$ and
$T = 170$ MeV, $y_{Tmax} = 0.50$ are essentially identical. 
  
\begin{figure}[hbt]
\vskip 1.0truecm
\epsfig{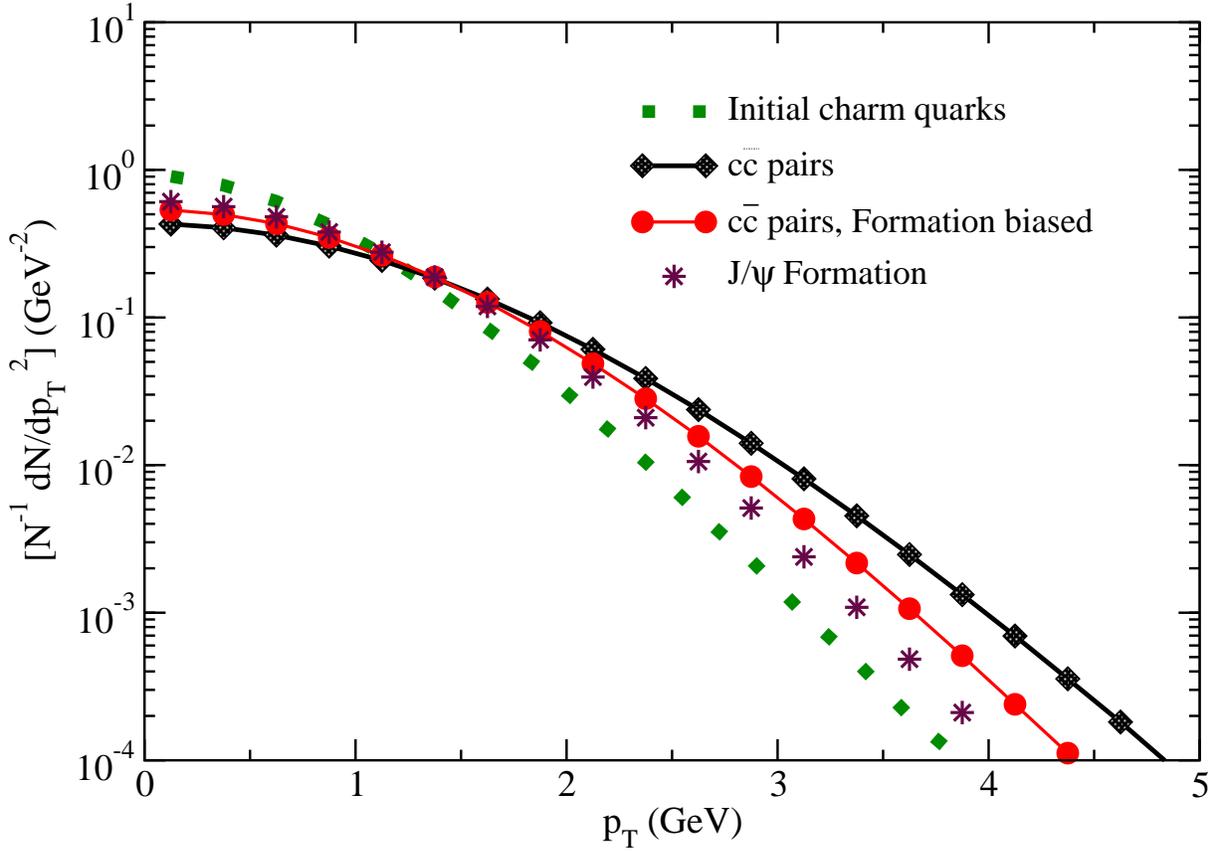}
\caption{(Color online) Transverse momentum spectra comparison for $\J$
and $ccbar$ pairs for a thermal plus flow initial quark
distribution.
}
\label{blastjpsiandccbarpt}
\end{figure}
\begin{figure}[hbt]
\vskip 1.0truecm
\epsfig{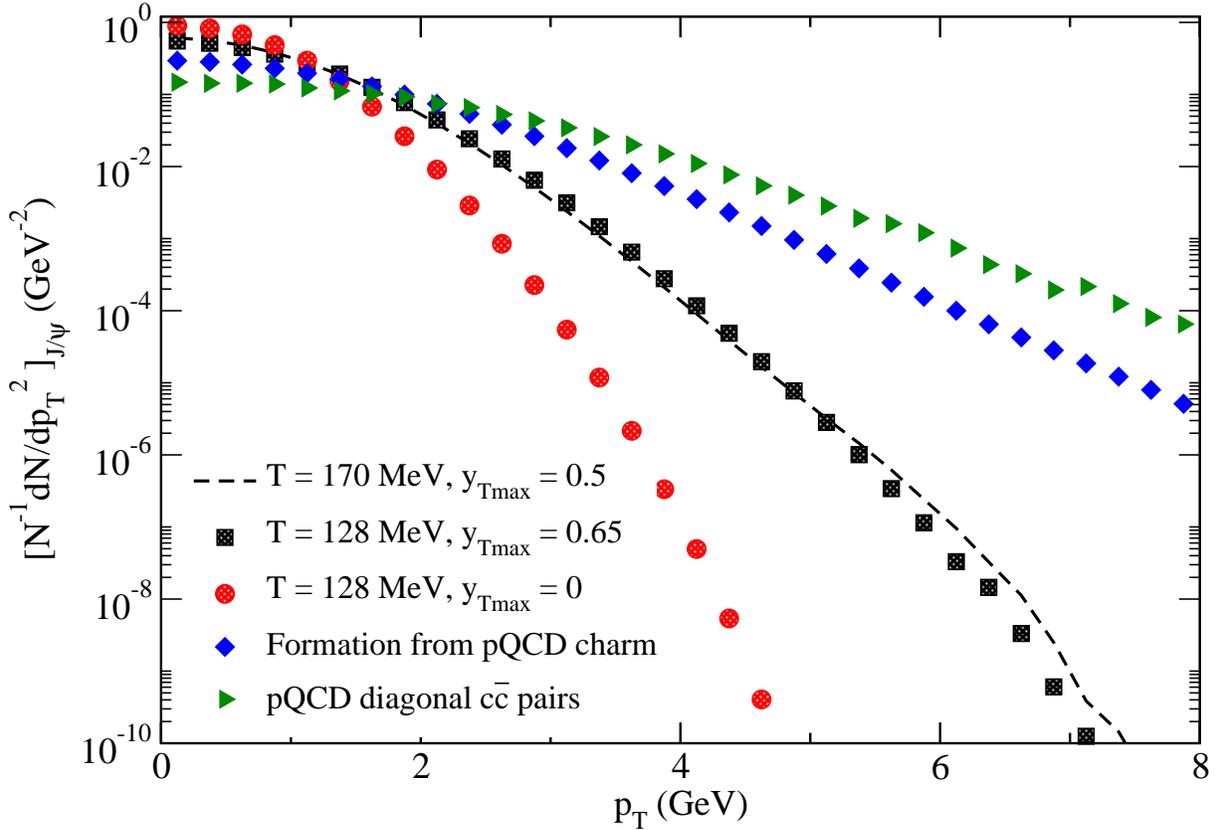}
\caption{(Color online) $\J$ formation transverse momentum spectra
dependence on initial charm quark distributions.
}
\label{jpsiformblastandpqcd}
\end{figure}

We show in Fig.~\ref{blastjpsiandccbarpt} 
the transverse momentum spectra evolution which
follows from charm quarks with $T = 170$ MeV, $y_{Tmax} = 0.50$.
One sees a downward progression in 
widths, starting with $\pts = 2.5~\gev^2$ for $\ccbar$ pairs, 
to $2.0~\gev^2$ for weighted $\ccbar$ pairs 
and $1.7~\gev^2$ for $\J$ formation.  
{\it It is important to note that the concept of diagonal pairs no longer 
exists for this set of charm quarks, since their formation identity has
been entirely erased during the thermalization process.}   
As might have been expected with a thermal distribution, these widths are
all smaller than any of those which follow from pQCD charm quark distributions.
This is shown in Fig.~\ref{jpsiformblastandpqcd}. 
The filled diamonds and
triangles are the $\J$ formation and diagonal $\ccbar$ pair distributions
which follow from the pQCD quark distributions as calculated in 
Sec. \ref{pqcd}.  The solid squares and dashed line are for the two
equivalent thermal plus flow charm quark distributions, and for completeness
a formation spectra without charm quark transverse flow is shown by the
solid circles.  It is clear that in-medium formation of $\J$ starting from either
the pQCD or thermal plus flow charm quark distributions predict 
transverse momentum widths (characterized by $\pts$) which are markedly smaller
than that for initially-produced $\J$. The difference between the two
formation possibilities would be reflected by the significant difference 
between the shapes of the $p_T$ distributions.  

\begin{figure}[hbt]
\vskip 1.0truecm
\epsfig{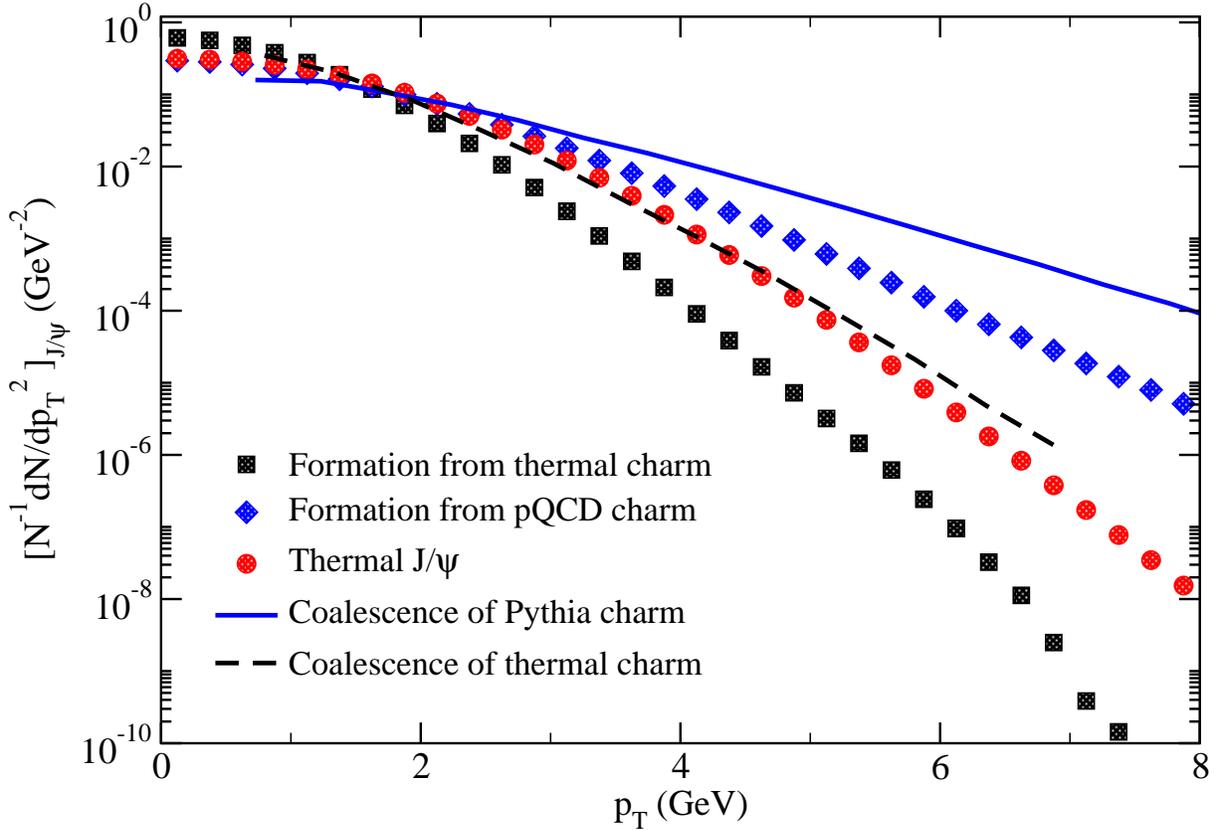}
\caption{(Color online) $\J$ formation transverse momentum spectra
comparison with alternative models.
}
\label{jpsimodelcomparison}
\end{figure}

In Fig. \ref{jpsimodelcomparison} we compare these results with 
alternate model calculations in the literature.  
The filled diamonds and squares are the $\J$ transverse momentum
spectra from our in-medium formation calculations using the two extreme
assumptions about the charm quark momentum distributions, 
pQCD (diamonds) and thermal plus flow (squares).  We compare this
with a calculation which uses a coalescence model to produce $\J$
from charm quark pairs 
\cite{Greco:2003vf}. The solid line is the result using charm 
spectra generated by Pythia (compare with our pQCD formation), and
the dashed line uses thermal plus flow charm spectra in the
coalescence model (compare also with our formation using thermal plus
flow).  One sees that the systematics of the model spectra are the same, but
the coalescence process yields somewhat broader spectra than our
direct in-medium formation.  Also shown by the filled circles is
the $\J$ spectrum which follows from a statistical model of hadronization.
Here the $\J$ would inherit a thermal plus flow kinetic distribution
\cite{Bugaev:2002fd}.  This distribution is broader than the underlying
charm quark thermal plus flow case, but narrower than the pQCD formation
and coalescence processes.  The width of the $\J$ transverse momentum
spectra at RHIC has also been predicted in \cite{Zhu:2004nw}.  The
value quoted for the most central collisions 
is $\pts$= 3.3 $GeV^2$ for the most central
collisions.  For comparison, our pQCD formation process predicts
3.6 $GeV^2$ in a region where it is the dominant process.


\section{Summary}
We have shown that in-medium formation of heavy quarkonium states
utilizing charm and anticharm quarks coming from independent hard
scatterings results in normalized momentum
spectra which retain a memory of the underlying quark distributions.
In the case that the quark distributions follow from initial production
in pQCD processes, both the rapidity and transverse momentum spectra
of the formed heavy quarkonium will be narrower than those expected from
diagonal pairs in the absence of a color-deconfined medium.  Explicit
calculations for $\J$ formation in Au-Au collisions 
at 200 GeV are performed, using
initial data in $pp$ and d-Au interactions to fix some parameters.
A striking feature is the non-monotonic behavior of the transverse
momentum spectrum widths $\pts$ in the progression $pp$ to pA to AA.  
In the absence of in-medium formation, one would expect $\pts$ to 
increase monatonically with the colliding system size, due to the initial
state confinement and nuclear broadening effects.  Our calculations for
in-medium formation in AA collisions predict $\pts$ substantially
smaller than expected if the $\J$ were due to initial production 
(see Fig.~\ref{ptpredictions}.)  In fact, the predicted 
$\pts_{AA}$ = 3.59 $GeV^2$ is
even smaller than the measured value in d-Au interactions, 
$\pts_{d-Au}$ = 4.25 $GeV^2$.

This 
effect is maximum  for central AA collisions, and will revert to
the ``normal" behavior for peripheral collisions. Uncorrelated pairs,
composed of one quark and one antiquark which do not originate from
the same initial interaction, 
have an invariant-mass spectrum which
becomes harder for larger $p_T$ of the pair. Given that the
charmonium formation dynamics favors lighter invariant masses, we
expect that the contribution of uncorrelated pairs will be more
dominant at smaller $p_T$, leaving a signature in the overall $p_T$
slope. The softening of the $p_T$ spectrum for uncorrelated pairs, in
other words, is the result of an increased contribution to charmonium
production at lower $p_T$.

  We have also
considered the formation process using thermal charm quarks which
flow with the expanding medium.  The resulting spectra are 
substantially narrower and retain a form which reflects that of
the underlying heavy quarks.  Overall, our normalized spectra appear to be quite
robust with respect to variations of the model parameters, independent of 
the absolute magnitude of in-medium formation.


\acknowledgments

This research was partially supported by the U.S. Department of
Energy under Grant No. DE-FG02-04ER41318.



\end{document}